# Records of Sunspots and Aurora Candidates in the Chinese Official Histories of the *Yuán* and *Míng* Dynasties during 1261-1644

Hisashi Hayakawa (1-3), Harufumi Tamazawa (4), Yusuke Ebihara (5, 6), Hiroko Miyahara (7), Akito Davis Kawamura (4), Tadanobu Aoyama (3), Hiroaki Isobe (6, 8)

(1) Graduate School of Letters, Osaka University, Toyonaka, Japan
(2) Research Fellowship for Young Scientists of Japan Society for the Promotion of Science, Tokyo, Japan
(3) Graduate School of Letters, Kyoto University, Kyoto, Japan
(4) Kwasan Observatory, Kyoto University, Kyoto, Japan
(5) Unit of Synergetic Studies for Space, Kyoto University, Kyoto, Japan
(6) Research Institute for Sustainable Humanosphere, Kyoto University, Kyoto, Japan
(7) Musashi Art University, Tokyo, Japan
(8) Graduate School of Advanced Integrated Studies in Human Survivability, Kyoto University, Kyoto, Japan

**Abstract**

Records of observations of sunspots and auroras in pre-telescopic historical documents provide useful information about past solar activity both in long-term trends and short-term space weather events. In this study, we present the results of a comprehensive survey of the records of sunspots and aurora candidates in the *Yuánshī* and *Míngshī*, Chinese Official Histories spanning 1261−1368 and 1368−1644, based on continuous observations with well-formatted reportds conducted by contemporary professional astronomers. We then provide a brief comparison of these data with Total Solar Irradiance (TSI) as an indicator of the solar activity during the corresponding periods to show significant active phases between 1350s-80s and 1610s-30s. We then compared the former with contemporary Russian reports for naked-eye sunspots and the latter with contemporary sunspot drawings based on Western telescopic observations. Especially some of the latter are consistent with nitrate signals preserved in ice cores. These results show us some insights on not only minima and maxima of solar activity during 13[th]-17[th] century.

**Introduction**

It is important to understand how extreme space weather events can cause significant social and economic impacts in addition to their impacts on solar and stellar physics. The most intense geomagnetic storm in history of ground-based telescopic observations is believed to be so-called the "Carrington event" in 1859 (Carrington 1859; Kimball 1960; Tsurutani et al. 2003; Cliver & Svalgaard 2004; Hayakawa et al. 2016b), with an estimated disturbance storm time (Dst) index of ∼ −1760 nT (Tsurutani et al. 2003). In this event, a white light flare within a great sunspot on 1859.09.01 observed by Carrington (1859; 1863) caused a series of severe magnetic storms and low-latitude auroras up to 23° in magnetic latitude such as Hawaii, Caribbean Islands, or Southern Japan were observed (Kimball 1960; Cliver & Svalgaard 2004; Hayakawa et al. 2016b). Recent studies warn us that the same-scaled solar storms can cause severe disasters as serious as 2 trillion USD in case they hit the modern civilization (National Research Council 2008), so that extreme events like Carrington event require intensive studies. Even with the centuries of history of the telescopic observations for sunspots since 1611 (Hoyt & Scahtten 1998; Owens 2013; Clette et al. 2014; Svalgaard & Schatten 2016) and solar flare observation since the Carrington event, the history of our telescopic observation can be short for discussions on long-term solar activity and extreme space weather.

There is possibility that even more extreme space weather events occurred in the pre-telescopic era. Maehara et al. (2012) reported their discovery of superflares on solar type stars (G-type stars). The energies of those superflares are considered to be 10-1000 times as large as solar flares captured in the modern telescopic observations. Notsu et al. (2015a, 2015b) reported that some of those stars with superflares have large starspots and a relatively low rotation velocity, which supports the hypothesis that there is possibility that superflares might occur in our present Sun (Shibata et al. 2013), although it is still in controversy (e.g., Schäfer et al. 2000; Auranier et al. 2013; Candelaresi et al. 2014; Nogami et al. 2014; Honda et al. 2015). Miyake et al. (2012, 2013) discovered an anomalous increase of atmospheric carbon-14 in tree rings in 774/775 and in 993/994. These anomalous increases of carbon-14 would give a sign of an increase in cosmic ray fluxes during those periods. While various ideas of the origin of these events have been suggested, Mekhaldi et al. (2015) and Miyake et al. (2015) detected signals of beryllium-10 in ice cores as well to relate these signals with the solar energetic particles (SEPs) caused by extreme solar flares. In order to identify the origin of such cosmic ray events, survey of auroras and sunspots in historical documents have been conducted (e.g. Allen 2012; Usoskin et al. 2013; Stephenson 2015; Hayakawa et al. 2017a; Tamazawa et al. 2017).

Indirect data would provide important information about the long-term solar activity and the past space weather events. One is with proxy records; for example, nitrate concentration in ice cores as an index of SEPs event (McCracken et al. 2001), and cosmogenic radionuclides such as carbon-14 and beryllium-10 as proxies for both long-term solar activity (Solanki et al. 2004; Steinhilber et al. 2009) and extreme space weather events (Miyake et al. 2012, 2013, 2015; Mekhaldi et al. 2015). The other is with records of low latitude aurora and naked-eye sunspots in historical documents (Shiokawa et al. 2005; Vaquero & Vázquez 2009; Usoskin 2013). With a larger sunspot appearance, a larger flare is possible (Shibata et al. 2013); then a larger flare can cause more intense geomagnetic storm as well as aurora in lower latitude (Odenwald 2015; Takahashi et al. 2016; Takahashi & Shibata 2017). Historical documents provide relatively precise information on observational dates and sites, while radioisotope proxies can only provide yearly resolution in general (e.g. Keimatsu 1970; Vaquero & Vázquez 2009; Hayakawa et al. 2015, 2017a). Therefore, discussing about historical records on naked-eye sunspots and low-latitude auroras can provide more detailed insights and information on past solar activity, and hence attracted several authors to the original historical documents written in various areas and languages[1].

---

[1] The significant studies based on original historical documents in each region are as following: China (Keimatsu 1970, 1971, 1972, 1973, 1974, 1975, 1976, the Beijing Observatory 1988; Clark & Stephenson 1978; Yau & Stephenson 1988; Yau et al. 1995; Willis & Stephenson 1999; Xu et al. 2000; Hayakawa et al. 2015, 2016a; Kawamura et al. 2016; Tamazawa et al. 2017), Korea (Lee et al. 2004), Japan (Kanda 1933; Matsushita 1956; Nakazawa et al. 2004), Babylonia (Stephenson et al. 2004; Hayakawa et al. 2016c), West Asia (Basurah 2006; Hayakawa et al. 2016d), Europe (Fritz 1873; Link 1962; Dell'Dall'Olmo 1979; Stothers 1979; Vaquero & Trigo 2005; Vaquero et al. 2010), Russia (Vyssotsky 1949), the Tropical Atlantic Ocean (Vázquez & Vaquero 2010), and so on.

This study aims at re-surveying and compiling the Chinese records of observations of naked-eye sunspots and low-latitude auroras in 13[th]-17[th] century, with explicit criteria of the selection of the source documents and the keywords used in the survey, to find out candidates of extreme solar events and to compare them with scientific datasets. Therefore, we present the data from the *Yuánshǐ* (元史) and the *Míngshǐ* (明史), the Official Histories in the era of *Yuán* and *Míng*. We will refer to data from the *Yuánshǐ* (元史, hereafter YS) and to those from the *Míngshǐ* (明史, hereafter MS).

## Historical Background of the Eras of *Yuán* and *Míng*

The target era, in the *Yuán* (元) and *Míng* (明) dynasties during 1261[2]−1368 and 1368−1644, witnessed significant changes of Eurasian history both in the perspectives of history and climate change. Historically, in this era, the Mongolians led by Genghis Khan swept over all of Eurasia and established four khanates, including the *Yuán* dynasty in China. They started vast communications between the East and the West under their rule, the so-called "Pax Mongolica," and caused a great change in the political systems across Eurasian regions (D'Ohsson 1834; Spuler 1939; Spuler 1943; Boyle 1977; Sugiyama 2004), exposed Chinese astronomy under influence of Islamic astronomy as described later. In the latter half of 14th century, the Mongolian Khanates lost their hegemony in Eurasia. In China, the *Yuán* dynasty was weakened by the Red Turban Rebellion caused by a series of famines, and the *Míng* dynasty took the hegemony on the mainland of China to drive the Mongolians away. The *Míng* dynasty turned back to Chinese traditional policies in every domain, including astronomy, although some of the astronomical policies in the previous *Yuán* dynasty were preserved[3] (Yabuuchi 1967).

Not only history, but also climate and solar activity have changed dramatically in this era. The era witnessed the finish of Medieval Warm Period (MWP: 1000-1250) and the start of Little Ice Age (LIA). From the viewpoint of solar activity, this era is contemporary with two grand minima: the Wolf minimum (1280-1350) and Spörer minimum (1450-1550) (Eddy 1977a, 1977b; Usoskin et al. 2007). It is known that solar activity can influence the terrestrial climate, although the detailed mechanisms of solar impact on climate are not clear (Gray et al. 2010). Therefore, it is important to survey records in this era also to understand solar-terrestrial relation.

## Methods

### Astronomical observations in the *Yuán* dynasty and *Míng* dynasty

Despite the changes which Chinese astronomy experienced in these ages, Chinese astronomers kept the same formats for recording celestial events. Therefore, Chinese astronomical records after the dynasties of *Yuán* and *Míng* also kept their remarkable consistency to be used as scientific data among historical records worldwide, as mentioned by Keimatsu (1976) and Hayakawa et al. (2015). Their records are the results of observations by professional astronomers with continuous observations at specified locations, which show us exact dates of the astronomical phenomena in question, often with detailed notes such as motions, shapes, and colors. In these eras, too, their main purpose was "astro-omenology" for policy makers (Pankenier 2013). This purpose can be observed

---

[2]  The name of *Yuán* started to be used in 1271 formally under Khubilai Khan. However, after conquering *Běijīng* to drive *Jīn* dynasty away, the Mongolian started to use the observatory at *Běijīng* and recorded astronomical events in their records. We explain this later (D'Ohsson 1834; Sugiyama 2004).
[3]  For example we can confirm that Muslim astronomers were working in the early age of this dynasty as well (MS, Calendar I, 516-17; MS, Staffs III, 1811)

in contemporary publications for astro-omenology, such as *Tiānyuán Yùlì Xiángyìfù* (天元玉曆祥異賦; TYX).

The motivations for observations during these ages were not changed at all compared with those of other dynasties. In traditional Chinese thought, the astronomical phenomena were thought to be signs from the heavens to the emperors reflecting their politics (YS, Astronomy, 989-990). as was also the case with the *Sòng* (宋) era (Hayakawa et al. 2015). Additionally, in YS, such records occupied the treatises of five elements, with the aim of compiling the records that show upsets of the balance of the five elements (五行): wood, metal, fire, water, and soil. Therefore, in order to deliver those celestial messages to emperors, Chinese dynasties had founded observatories near their capital cities, e.g., in *Sòng* dynasty (Hayakawa et al. 2015). Just after the Mongolians conquered *Yàn* (燕, i.e., modern *Běijīng*), a capital city of the previous *Jīn* (金) dynasty[4], the Mongolians founded a new dynasty, *Yuán*, and established the same governmental facilities including observatories (YS, Astronomy, 989-990).

However, the Mongolian authorities under Kublai Khan started innovation in Chinese astronomy. In 1271, they brought *sèmùrén* (色目人), such as Jamāl al-Dīn (札馬剌丁), and founded an Islamic Observatory (回回司天臺) at their capital *Dàdōu* (大都), the present *Běijīng*, to observe celestial events and revise calendars (YS, Staffs VI, 2297). They brought the latest astronomy from western Asia and caused a considerable exchange of astronomical technology between western Asia and China. (Rufus 1939; Yabuuchi 1967; Vesel 2002; Isahaya 2010)

In 1279, a huge change was brought into Chinese astronomy, and they started to offer more observation points in their records. A well-known astronomer, *Guō Shǒujìng* (郭守敬)[5], requested that Emperor Kublai Khan establish 27 observatories all over the *Yuán* dynasty territory (cf. Table 1; YS, *Guō Shǒujìng*, 3848). These observatories are located all over the Mongolian territory from 15° N to 65° N. From that time, the Chinese Official Histories, including YS, involved records of astronomical observations from various observatories, though regional records are not only from cities with observatories.

*Míng* dynasty also followed this policy and founded observatories, not only in their capital cities, but also in their local cities just as did the following *Qīng* (清) dynasty (Kawamura et al. 2016)[6]. It was also in the late *Míng* era that Chinese astronomers made contact with the Western modern sciences brought by Western missionaries, such as Matteo Ricci, for the first time.

**Source Documentation**

In this study, we examine records such as those in the treatises of astronomy (天文志), the treatises of five elements (五行志), and imperial chronicles (本紀) in contemporary Official Histories, i.e., the *Yuánshī* (YS: 元史) and the *Míngshī* (MS: 明史), to investigate the records of sunspots and aurora candidates in 1261-1368 and 1368-1644. These ages (1261-1644) overlap with the beginning of the so-called LIA. Their bibliography is as follows:

**Contemporary Official Histories**

YS: Sòng Lián, *Yuánshī* (元史), I-XV, Běijīng: Zhōnghuá Shūjú, 1976 [critical editions in Chinese].

---

[4] A Khitai-origin dynasty ruled the northern half of China from 1127 to 1234.

[5] An astronomer lived during 1231 and 1316. He is quite a well–known astronomer who revised the old Chinese calendars and established a new calendar called *Shòushílì* (授時曆) with the influence of contemporary Arabic calendars. This calendar is one of the most precise historical solar calendars and was in use up to CE 1644.

[6] Keimatsu (1976) has already mentioned that local observatories were installed after the *Míng* dynasty, but it seems he did not notice the situation for *Yuán*.

MS: Zhāng Tíngyù, *Míngshǐ* (明史), I-XXVIII, Běijīng: Zhōnghuá Shūjú, 1976 [critical editions in Chinese].

**Historical Documents for Additional Discussion**

TYX: *Tiānyuán Yùlì Xiángyìfù* (天元玉曆祥異賦), a manuscript at 305-257, Naikaku Bunko, Books of Shoheizaka Gakumonjo, in the National Archives of Japan (国立公文書館 昌平坂學問所本 內閣文庫 305-257) [a manuscript in Chinese]

SS: Tuōtuō, *Sòngshǐ* (宋史), I-XL, Běijīng: Zhōnghuá Shūjú, 1977 [critical editions in Chinese].

*Jiùtángshū*: Liú Xù, *Jiùtángshū* (舊唐書), I-XVI, Běijīng: Zhōnghuá Shūjú, 1975 [critical editions in Chinese].

*Qīngshǐgǎo*: Zhào Ěrxùn, *Qīngshǐgǎo* (清史稿), I-XLVIII, Běijīng: Zhōnghuá Shūjú, 1976 [critical editions in Chinese].

ПСРЛ: *Полное собрание русских летописей*, I-XLIII, 1846-2004 [critical editions in Russian]

In contrast to previous catalogs for contemporary aurora observations in China compiled by Keimatsu (1970, 1971, 1972, 1973, 1974, 1975, 1976), the Beijing Observatory (1988), Yau et al. (1988, 1995), or Xu et al. (2000), we excluded other historical documents such as Official Imperial Records (実録) or Local Treatises (地方志) to limit our survey to contemporary Official Histories to avoid apparent increase of records, according to the following reasons:

First, we must note that the Official Histories were compiled from the original daily records written by the observers and historians that had seem to be lost. The astronomical records were not an exception. The author of MS mentions in the Astronomical Treatise explicitly that the astronomical data in MS were selected from the original observations (MS, Astronomy III, 417). The process for the Official Histories to be compiled is as follows: Every Chinese authority had Departments of History (太史局) to record historical matters, including astronomical records from the observatories. Every year, these records were compiled into an Official Yearly Record (起居注). When a contemporary emperor died, the Chinese Official Historians edited these Yearly Records to make an Official Imperial Record (実録) for every emperor. When a dynasty perished and was replaced by newcomers, these new comers edited the Official Imperial Records of the previous dynasty to compile an Official History, in order to authorize their succession from the previous dynasty (Takeuchi 2002). This tradition in China started as early as the *Táng* dynasty (*Jiùtángshū*, Staffs II, 1845; *Jiùtángshū*, Staffs II, 1853). The Official Imperial Records are still available for the eras of the *Míng* dynasty, and the Official Yearly Records are still available from the era of the *Qīng* dynasty, although those before the *Yuán* dynasty had already been lost and unavailable.

Second, the Local Treatises (地方志) started to increase their number considerably in these eras. The contemporary local intellectuals started to compile books of history and geography for their homelands in these ages, including records for astronomical events in chapters regarding omens (祥異). Almost all of the Local Treatises that are still available were written in the eras after the *Míng* dynasty (Beijing Observatory 1985) that is consistent with the contemporary development of printing technology brought the media revolution (Oki 2009)

Although extending our survey to these records can bring us significantly more records for aurora candidates and sunspots than those from the Official Histories, it can easily overestimate the actual amount of observations. In other words, including other kinds of records without any criteria can make us overrate or underrate solar activity by the bias how many historical sources are still available today for each period. For example, Keimatsu (1976) seemed to overestimate the solar activity in the 16th century as the above-mentioned biases are not taken into account. Therefore, we limited our surveys for records of aurora candidates or sunspots on Official Histories for further comparison with scientific data to avoid contamination derived from amounts of historical sources.

**Target Terms**

YS and MS include descriptions of considerable phenomena observed in the sky. Since we are interested in the past solar activities, we surveyed descriptions that could be regarded as records of sunspots or aurora candidates. For this purpose, we used a search engine, Scripta Sinica (新漢籍全文; http://hanchi.ihp.cinica.edu.tw) provided by Academia Sinica in Taiwan that incorporates all the text data of the Official Histories. This allowed us to automatically flag sentences that included keywords such as "black spot (黒子)" and "red vapor (赤氣)," which may refer to sunspots and red auroras, respectively. Once the sentences that include the keywords were flagged, we cross-checked the resultant data with published critical editions. We also calculated the moon phase to determine the sky conditions for each date of the observations.

**Sunspot records**

Sunspots are described as black spots (黒子) or black vapors (黒氣) in the sun (Saito & Ozawa 1992; Hayakawa et al. 2015). TYX, a contemporary manual for astro-omenological divinations, includes vivid drawings for sunspot record with descriptions as is shown in Figure 1. On the other hand, we did not include those "near the sun (日傍/日旁)" as they are not in the sun and hence are not sunspots as explained in Appendix I. Although we could not find such records in YS, we found some relevant records in MS. There are several possible reasons for the difference in how these Official Histories treated the sunspot records. The first is the astronomers in the *Yuán* Dynasty simply did not observe the sun as much as those in other dynasties. The second is the solar activity itself, as the period of the *Yuán* Dynasty overlaps the Wolf minimum (1280-1350). Comparing the sunspot records in the SS with YS, we found one sunspot candidate in 1276 during the period when both dynasties of *Yuán* and *Sòng* coexisted (CE 1261-1279), while SS has 38 sunspot candidates in their whole record in total (Hayakawa et al. 2015). These results suggest naked-eye sunspot can have been observed in the end of the MWP in this early *Yuán* era, although YS does not include sunspot records.

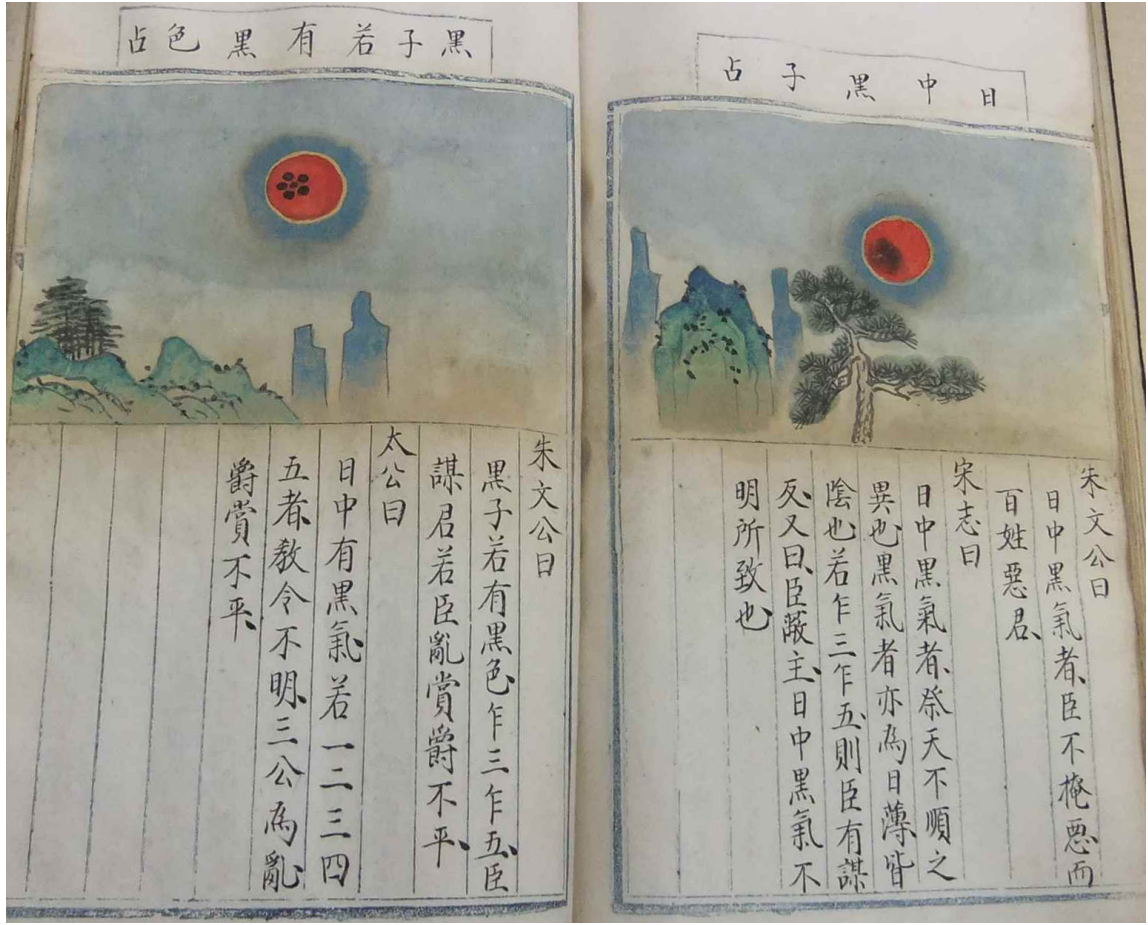

Figure 1: drawings for sunspots in TYX (I, 39b-40a)

**Records of aurora candidates**

Although the ancient Chinese did not seem to know the physical nature of the phenomena, there are a number of records that can be considered as the observations of auroras (Keimatsu 1969a, 1969b; Saito & Ozawa 1992). TYX also includes vivid drawings with descriptions as is shown in Figure 2

As in our previous survey for SS (Hayakawa et al. 2015), we assumed that the records of luminous phenomena observed at night are potentially those of auroras. In addition, we accounted for terms that were newly suggested by Hayakawa et al. (2016a) and supported by Carrasco et al. (2017). Therefore, we surveyed the words that refer to luminous phenomena: "vapor (氣)," "light (光)," "cloud (雲)," and "white/unusual rainbow (白虹/虹蜺)" in YS and MS.

From the list of the potential aurora candidates, we manually removed two types. The first are those without dates. Most of these are found in the chapters explaining how astro-omenology is performed, which includes conversations between the emperor and servants or sages without date of observations. That is because they are unlikely to be the direct records of the observations. The second type includes those seen during the daytime, although we leave records observed in the twilight. Some of the daytime phenomena are presumably the solar halo. Nevertheless, we did not limit our survey on those written as observed "at night" explicitly. As we show following, we have examples of records for the same record in YS (see, YS#A10), one without observational time and another with observational time[7].

    Records of aurora candidates often include information about their color, motion, and direction; the length, shape, and the number of their bands; and sometimes the location of the observation when it was not made in the capital city.

    The color is described as white or red, though such records were described in five colors or their mixture in other eras (Hayakawa et al. 2015; Kawamura et al. 2016; Tamazawa et al. 2017). In the traditional Chinese thought, called the *Wǔxíng* (五行) or Five Elements, the world consist of five elements, i.e., metal, fire, wood, soil, and water, which correspond with the colors of white, red, blue, yellow, and black, respectively (Hayakawa et al. 2015)[8]. Their motions and directions are usually given by the eight points of the compass. The shape and number of bands are described only in some of the recorded events. The shapes are expressed in a figurative way, such as "like serpents," "like fire," or "like blood."

    Occasionally, these records involve information about the constellations, the planets, or the moon that accompany the auroras. Their lengths are given in the units of "*chǐ*," "*zhàng*," or "*lǐ*." According to Tonami et al. (2006), these units have been converted into the modern units as 1 *chǐ* equal to 30.72 cm in the *Yuán* era and 31.10 cm in the *Míng* era, 1 *zhàng* equal to 10 *chǐ* in both eras, and 1 *lǐ* equal to 553 m in the *Yuán* era and 560 m in the *Míng* era. Currently, we do not know how these explanations of length represented the actual distribution of the events seen in the sky (Hayakawa et al. 2015).

---

[7] Note that the similar cases are seen not only in YS but also other Chinese Official Histories as well. Another good example is found in SS for the aurora record on 1138.10.06 (SS, Five Elements IIb, p1412; SS, Astronomy XIII, p1314).

[8] There are mistakes in Hayakawa et al. (2015), which included "green" instead of "blue" in the 5 colors.

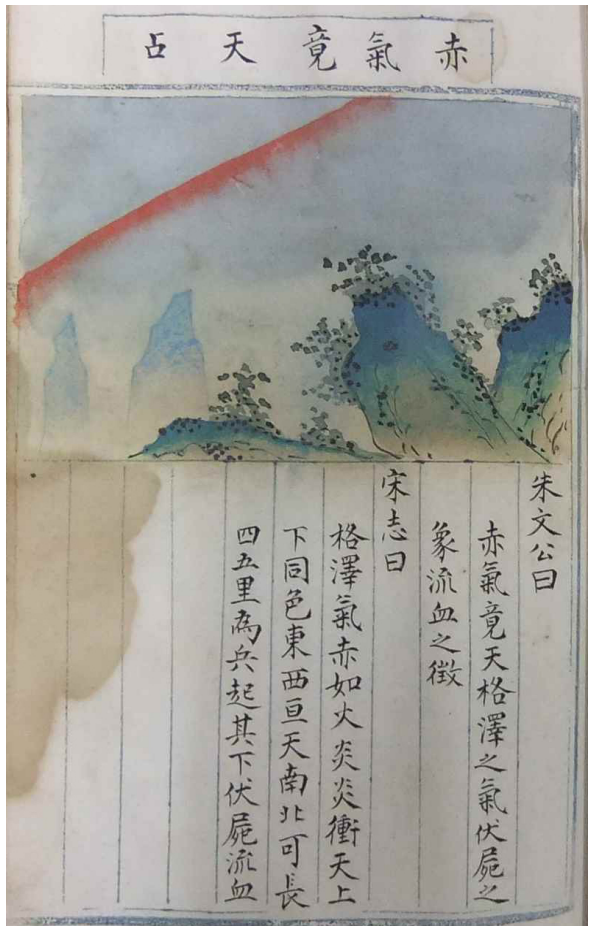

赤氣竟天占

朱文公曰
赤氣竟天格澤之氣伏屍之
象流血之徵

宋志曰
格澤氣赤如火炎炎衝天上
下同邑東西旦天南北可長
四五里為兵起其下伏屍流血

Figure 2: red vapor in TYX (VI: 38a)

## 3. Results

   We found 20 aurora candidates in YS and 10 aurora candidates and 26 sunspot candidates in MS, as shown in Appendix II. The format of the description is as follows: the ID number, the date of observation (year. month. date), the original texts in classical Chinese, the reference for the original texts, and the translation into English. The details of the each record with the original text and the bibliographic information are available at our website (http://www.kwasan.kyoto-u.ac.jp/~palaeo) as well. The date in Chinese lunar calendar is converted to that of the Julian calendar up to 1582 and to the Gregorian calendar from then forth, based on chronological tables by Wang et al. (2006). These overall results are summarized in Table 1. Figure 3 shows the annual counts of the said records.

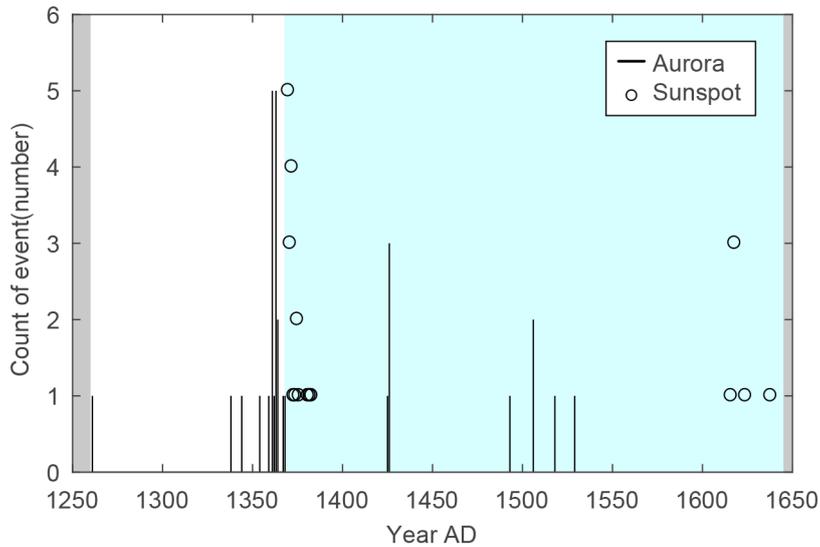

Figure 3: Change in the number of candidate sunspots and auroras during 1261–1644. The bars show the number of aurora candidates. The dots show the number of sunspots. Note that records of sunspot and aurora candidates are from documents written in different eras; therefore, white back ground represents *Yuán* era and blue back ground represents *Míng* era.

Some of the records we listed in Appendix II are not clearly dated or easily doubted as non-aurora record. YS#A1 was given in two records in the same sentence, Yau et al. (1995) and Xu et al. (2000), with its date as 1262.02.09 mistakenly, and we should correct this to the same date in 1261 according to Wang et al. (2006). YS#A10 is a good example of omitting observational time in the editorial process; comparing the sentences from *Shùndì* and Five Elements, we can immediately infer that they are from the same source by their dates, observational sites, and contents, but also find the difference in the observational time missing in the sentence from *Shùndì*, indicating its extra edit process from that from Five Elements. MS#A10 could be considered a comet by its duration (by applying the rule used by Kawamura et al 2016), and this is supported by the description for C/1618V1 in Kronk (1999) and Neuhäuser & Neuhäuser (2016). The information of those records is complemented, and the records are maintained on the list regardless of their degrees of reliabilities in order to show our complete result of survey.

## 4. Discussion

### Description of records as auroras

In this section, we discuss on some notable records from our list. The records only involving red or white colors are not the target of this section because those records are more commonly found among other.

YS#A2 may have an indication of the aurora that was seen at high elevation angles. If the aurora is seen near magnetic zenith, it would be regarded as corona aurora, which is a kind of rayed auroras. Aurora rays tend to be aligned along geomagnetic field lines because incident electrons tend to move along a magnetic field line. Thus, an observer sees the rays that converge toward the magnetic zenith. YS#A7, YS#A8, and YS#A9 cluster within four days and possibly from the same active region. This kind of clustering observation of aurora was known in the time of the Carrington flare from 1859.08.28 to 1859.09.04 (e.g. Loomis 1860; Kimball 1960). YS#A11 was described "like a rope" and that may indicate thin and long auroras like a rope, although we must not exclude the possibility that this was a comet. YS#A13 may be regarded to as an aurora because simultaneous observation was made in Japan (Willis & Stephenson 1999; Hayakawa et al. 2016a). YS#A14 was observed at two different positions simultaneously in China at *Píngyang Lù* and *Jìnníng Lù*. Descriptions are similar with each other. The former records three bands of white vapor, and the latter records three of them in the red heaven. YS#A15 may indicate a Type-A aurora in which the red aurora dominates in the upper part and the green aurora dominates in the lower part. A Type-A aurora is frequently observed at mid-latitudes during large magnetic storms. Both the red (630.0 nm) and green (557.7 nm) colors are associated with the emission of atomic oxygen. The separation of the color is understood to be a simple height effect under the dominance of a low energy population of incident electrons. At altitudes below 150 km, the green aurora dominates the red one because of quenching effects, which may appear as a yellowish aurora by an observer because of a physiological blending of the red and green colors (Chamberlain 1961). The green aurora is known to move like a serpent. When the aurora is faint, an observer may see the green aurora as a yellowish or whitish one moving like a serpent with the naked eye.

The blue-white color in MS#A2 is reminiscent of a Type-F aurora in which long rays with blue or purple color dominate. This typically occurs when the sunlight enhances the abundant $N^{2+}$ in the upper atmosphere. The enhancement of $N^{2+}$ results in the intensification of the blue or purple auroras in the upper part of the rays. This condition is satisfied when the upper atmosphere is in sunlight whereas the ground is in darkness. Therefore, a Type-F aurora may be observable just after sunset or just before sunrise. MS#A2 may describe the Type-F aurora, but we cannot definitely identify because of uncertainty of observation time. A Type-F aurora may be observable just after sunset or just before sunrise, in particular, in the spring or autumn. MS#A3 may indicate a black aurora in which aurora emissions are absent within a relatively uniform background emission (Davis 1978). The black part shows distinct forms, including black patches, black rings, black arc segments, thin black arcs, and black vortex streets (Peticolas et al. 2002). The observer may see the black arc or filamentary arc that splits into two (Trondsen & Cogger 1997). As for MS#A6, the yellowish aurora may be the result of a physiological blending of the red and green auroras that are the emission from atomic oxygen (Chamberlain 1961). In MS#A7, red vapor was related with fire because a "red color" was traditionally regarded as the color of "fire" in pre-modern Chinese tradition. As for MS#A8, the lower border of an aurora curtain shows straight, hooked, meandering, fragmentary, and vague forms (Kaneda et al. 1968). Occasionally, a straight aurora curtain with a slight bend is found, which may look like a bent ruler. MS#A10 could be considered a comet by its duration (by applying the rule used by Kawamura et al 2016), and this is supported by the description for C/1618V1 in Kronk (1999) and Neuhäuser & Neuhäuser (2016).

**Aurora candidates and moon phase**

One may claim regarding the reliability of these kinds of records is associated with considerable contamination by atmospheric optics events such as halos or moon dogs, which may appear to be similar to auroras, especially for people observing them in earlier times. We must firstly note that Chinese astronomers and historians generally placed halos in independent sections of "halos (暈適)" or "solar/lunar omens (日變/月變)" (e.g., MS, Astronomy III, 413-416) to distinguish them with aurora-like records[9]. Nevertheless, we apply the method used in Kawamura

---

[9] The same tendency can be found in other Official Histories in other ages as well. For example, in *Xīntángshū* (新

et al. (2016), which employs the interactive data language (IDL) programs from the astronomy user IDL library of NASA/Goddard Space Flight Center (Landsman 1993) and our program developed for numerically determining the moon phase of the given date. The advantage on lunar phase analysis is based on the nature of the atmospheric optics requiring a significantly luminous source. Therefore, the records of luminous phenomena in the night sky during the phases around the new moon are considerably more reliable than those around the full moon, although we must note that auroras can be visible when they are bright enough and the angular distance between Moon and the aurora is large enough[10] (e.g. the aurora reported by Barnard 1910)[11].

The results of distribution of aurora candidates in comparison with moon phase are presented in Figure 4 (YS) and Figure 5 (MS) including the recorded colors. Here we separately present diagrams for YS and MS because of the uncertainty over the consistency of observational criteria of both dynasties. As a result of separate analysis, we cannot gain enough record numbers statistically able to discuss the shapes of the distribution over the lunar phases, and hence our discussion will be limited with lunar phases to case studies. YS#A16, YS#A19, and MS#A6 are found in the dark nights around the new moon (the normalized moon phase of ≥ 0.9 or < 0.1); therefore, these records are more probable as auroras with the limited contamination by atmospheric optics.

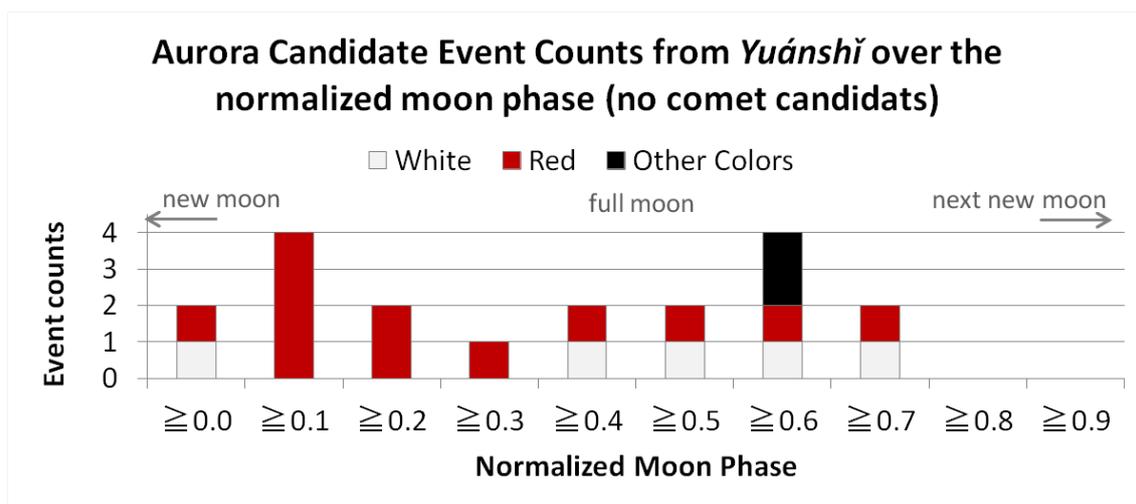

Figure 4: Number of aurora candidates from YS versus the normalized moon phase. No comet candidate with a long duration is found in the *Yuán* era.

唐書) as well, "white/unusual rainbows" were placed in sections of "solar halo," "lunar halo," and "unusual rainbows" to categorize their nature as shown in Hayakawa et al. (2016a). Ho & Needham (1959) examine *Jinshū* (晉書), one of Chinese Official Histories, to classify 26 technical terms for halos and claim that Chinese astronomers had already gotten deep knowledge on halo.

[10]   Although one may expect that it is very rare to see an aurora around full moon, the possibility depends on, at least, brightness of aurora, and angular distance between Moon and sky position (Krisciunas & Schaefer 1991). This means that the influence of the scattered moonlight is significantly low at large angle. According to calculation, Kenyon & Storey (2006) also claims that the scattered moonlight brightness can be lower than the brightness of aurora at certain zenith angle (Kenyon & Storey 2006). Because of these reasons, it is too speculative to rule out the possibility of aurora candidates around full moon to regard them as fogbow or night rainbow without detailed discussion on the view point of aurora science.

[11]   For example, Barnard (1910) states as follows: "September 4 (1908). 7h 30m: Bright aurora with nearly full moon. … 10h40m: The arch was very bright in spite of a bright moon." Thus, it is not so rare to see an aurora with a full moon.

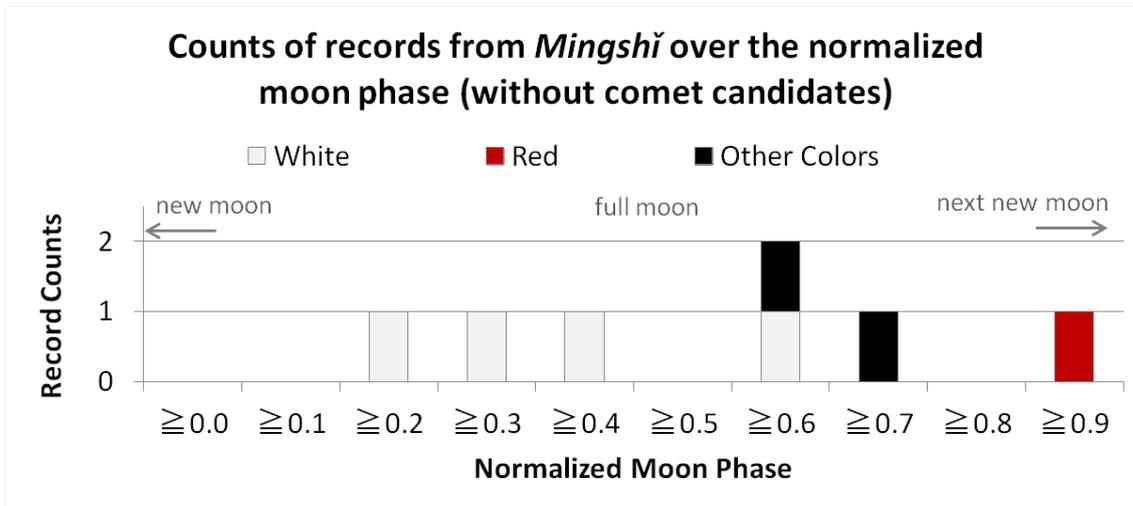

Figure 5: Number of aurora candidates from MS versus the normalized moon phase. Comet candidates with long durations are excluded for this graph.

**Naked-eye sunspots**

It is interesting that there are records of naked-eye sunspot after the Wolf minimum and the Spörer minimum, and a peak of records of aurora and naked-eye sunspots after the Wolf minimum. Previous studies (Heath 1994; Vaquero & Vázquez 2009; Hayakawa et al. 2015) suggested that the number of such large sunspots can be a good proxy of the overall solar activity. Indeed, the emergence of large sunspots that are visible to the naked eye is likely to be the necessary condition for the occurrence of extremely intense solar flares. Shibata et al. (2013) presented a scaling law that relates the energy of solar (and stellar) flares and the size of sunspots; for example, the maximum energy of a solar flare of a naked-eye sunspot can be around $10^{35}$ erg ($10^3$ times larger than modern scientific observation records). Therefore, trends of record numbers of naked-eye sunspots would be compatible with the theoretical suggestion of large flares associated with large sunspots and large magnetic energy.

The active phases of the solar activity between the Wolf minimum and the Spörer minimum were well attested to by contemporary naked-eye sunspots in other countries. They were observed in China, and also in Korea and in Russia. In Korea, we can find a cluster of sunspot observations during 1356–1378 (Lee et al. 2004). In Russia, we surveyed Russian medieval chronicles in Полное Собрание Русских Летописей (Complete Collection of Russian Chronicles, hereafter ПСРЛ)[12] to pick up sunspot records during 1360s and 1370s and cited into Appendix III in our article. We thus found 4 chronicles reporting on observations of enormously large sunspots in 1365 and 12 chronicles on those in 1371, while Vyssotsky (1949) had found 1 for the former (Appendix III, RC1) and 4 for the latter (Appendix III, RC5-7 and 9). We cite entries referring these sunspots from several Russian chronicles in Appendix III, RC1−RC4 for the sunspots in 1365 and RC5−RC16 for those in 1371, and add our translations to the original texts. A sunspot was recorded with the word "мѣста чрьны," with some orthographical variants, which means "black/dark spots in the sun". In Chinese Official Histories, while we could not find contemporary sunspot records in 1365, we found 3 contemporary sunspot records in 1371: MS#S6-S8. Especially, the sunspot recorded in MS#S7 in China is reported to last up to 30 days from 1371.06.13 to 1371.07.12. These records of sunspots in Russia seem written by accident; that is, observations by professional astronomers were not conducted continuously in Russia during the daytime; and, therefore, these records were very rare and very useful. It is under discussion how people observed naked–eye sunspots in those times. Although these entries of each year are very similar, we can't decide that they were written based on only one original entry. It is also possible that the same or different sunspots were observed from different places, and these observations were recorded in a fixed style. Vyssotsky (1949) has suggested a possibility that these sunspots were witnessed as the forests were burned due to severe drought and extreme hot weather when smokes and mists let people see the sunspot by their own eyes. In the Russian chronicles, it was reported that they could observe naked-eye sunspots because of the haze. These chronicles suggest that we should check weather conditions (including volcano eruption, fog, yellow sand, etc.) to determine if people could observe naked-eye sunspots. Interestingly, this peak between 1360s and 1370s overwraps with the period when "the Aral Sea was not existing" according to the contemporary geographical work by Ḥāfiẓ-i Abrū[13], a Tīmūrid historian. This may indicate that the intense solar maximum in this period had caused a severe warming in the central Asia as well. This is a topic for future discussion.

---

[12] ПСРЛ is a comprehensive collection of medieval Russian chronicles consisted of 43 volumes that cover all the chronicles up to those written in 18th century. This collection was entrusted to the Archaeographical Commission by the statute officially approved in 18 February 1837. At the same time, in order to accomplish this purpose, it was ordered to hand over all manuscripts of the Chronicles, kept in church and public libraries, under the control of the Commission (ПСРЛ: I, p. iii).

It is intriguing to compare historical sunspot records by Chinese traditional astronomy and "scientific" sunspot drawings by Western telescopic observations. It was after 1611 that the invention of the telescope allowed early modern western astronomers to start the telescopic observations (Gallilei 1613; Hoyt & Schatten 1998; Owens 2013). As MS has some records of sunspots dated after 1611 (MS#S21−S26), we tried a brief comparison between these datasets. We surveyed sunspot drawings in relevant periods within publications by early modern astronomers as we surveyed the published contemporary sunspot drawings by Scheiner (1630), Malapert (1633), and Gassendi (1658), which are conveniently reviewed by Zolotova & Ponyavin (2015) and Arlt et al. (2016). We found a probable coincidence between Chinese sunspot records and Western sunspot drawings in 1618.06/07, i.e. MS#S22 and MS#S23 and sunspot drawings by Malapert (1633: pp73-74) as shown in Figure 6. Malapert's relevant drawings cover the transitions of sunspots during 1618.06.21-29 and during 1618.07.07-19, while MS#S22 and MS#S23 fall on some dates during 1618.05.24-06.21 and during 1618.06.22-07.21, strictly speaking (Wang et al. 2006). Although its sunspot size is intriguing, Malapert (1633) seems not to draw shapes of sunspots but to place dots where sunspots locate, as long as seeing a series of his sunspot drawings (Malapert 1633: pp67-92).

---

[13]  Ḥāfiẓ Abrū 1997, *Juġrāfiyā-yi Ḥāfiẓ Abrū: muštamil bar juġrāfiyā-yi tārīḫī-i diyār-i ʿArab, Maghrib, Andalus, Miṣr, wa-Šām* (Tihrān: Intišārāt-i Bunyād-i Farhang-i Īrān).

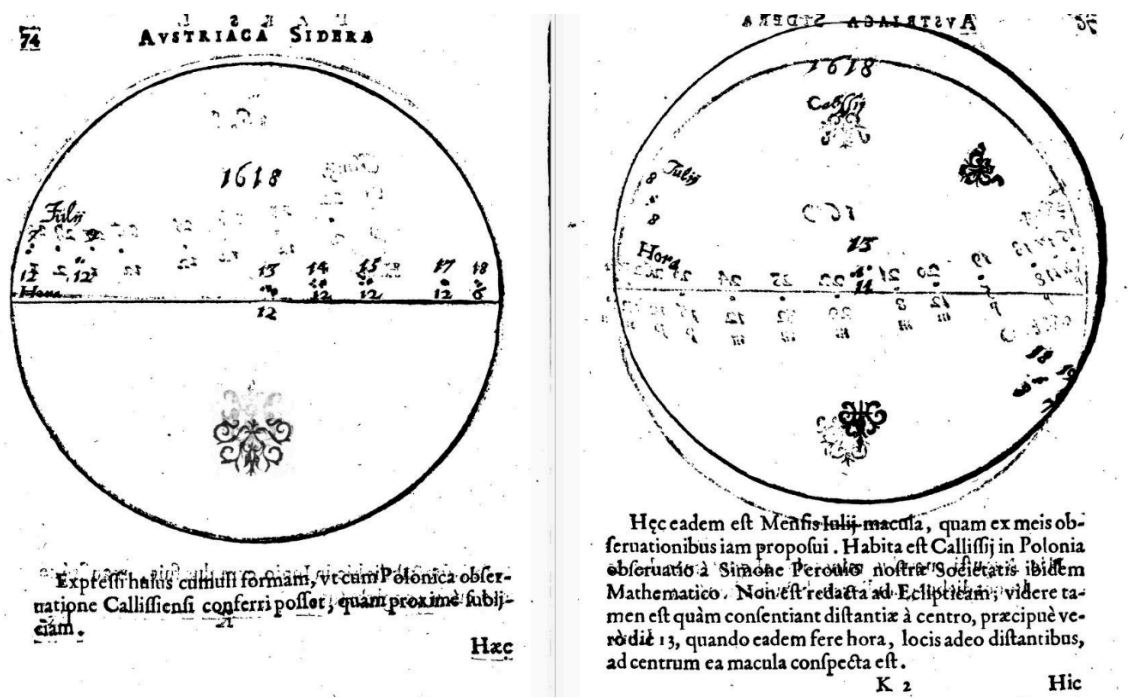

Figure 6: Sunspot Drawings by Malapert (1633: pp73-74) covering 1618/06/21-29 and 1618/07/07-19

Unfortunately, we could not find any more sunspot drawings with the same date in the sunspot records in MS and in the above-mentioned contemporary publications with sunspot drawings[14]. The closest drawings by date that we managed to find are as follows: on 1624.04.20-05.10 by Scheiner (1630: p217, Fig. 2) for MS#S22-S24, on 1624/05/17-26 by Scheiner (1630: p217, Fig. 6) for MS#S25, or in 1638/10-11 by Gassendi (1658, v4, p419) for MS#26. As Vaquero & Vázquez (2009) and Zolotova & Ponyavin (2015) mention several unpublished sunspot drawings in the early 17th century, further investigation may bring more sunspot drawings that can be compared with sunspot records in Official Histories.

**Comparison with reconstructed solar activity level and nitrate data**

In order to examine the correspondence of the above results with solar activity, we compared the timing of the data with the proxy-based solar activity level. Figure 7 (a)-(c) show the comparison with the anomaly of solar activity level in terms of total solar irradiance (TSI) reconstructed with multiple cosmogenic nuclide records (Steinhilber et al. 2009). Solar modulation parameter for cosmic rays was calculated based on the records of carbon-14 and beryllium-10, and was converted to TSI that reflects solar activity level. Both of the peaks in the number of naked-eye sunspot data coincide with the periods of high solar activity level at the end of the 14th century and the early 17th century that are just before the Spörer and the Maunder minimum, respectively. There seems to be a slight tendency for the naked-eye sunspot numbers to peak slightly after the solar activity maxima. The records of aurora candidates also show some correspondence with solar activity level (Figure 8).

---

[14]  MS#S24 is also in 1618 and hence Neuhäuser & Neuhäuser (2016) related this record with the left drawing of Figure 6 by Malapert during 1618.06.21-29. However, we need to note that Neuhäuser & Neuhäser (2016) interpret the misdated intercalary month of MS#S24 to intercalary fourth month to speculate this record on 1618.06.20-22. This interpretation is arbitrary and no more than speculation in a philological view point, otherwise supported by contemporary source documents such as *Míngshílù* (明実録). Even if their idea were correct, the date of *wùzǐ* (戊子) does not exist in intercalary 4[th] month in this year (Wang et al. 2006) and hence it is hardly supported. MS#S24 is totally different in description from MS#S22 that took place in "intercalary 4[th] month" and hence can hardly be regarded as from the same origin as well.

Our comparison shows some new insights to renew the comparison by Clark & Stephenson (1978, hereafter CS78) based on sunspot observations in Official Histories of dynasties in China and Korea and carbon-14 derived from Eddy (1976). While CS78 smooth carbon-14 history to place one of their peaks around 1400s (Figure 8 in CS78), the total solar irradiance reconstructed by Steinhilber et al. (2009) allow us to modify it around 1350s-80s (with its peak in early 1360s). CS78 also detect a series of sunspot observations in 1370s-80s as shown in our sunspot catalogue (Figure 7a, 7b; Table 3) as well. This peak is supplemented by aurora candidates in 1350s-60s recorded in YS as shown in our catalogue (Figure 8; Table 4). As explained previously, YS does not include sunspot records in it to show apparent absence of sunspots in 1350-1360s as shown in CS78. Hence, the comparison of records of sunspots with those of aurora candidates can resolve this difficulty and both TSI (C14 in CS78) and historical records show identical peaks in 1350s-1360s, between Wolf Minimum (1280-1350) and Spörer Minimum (1450-1550).

After Spörer Minimum, we can find two peaks around 1570s and around 1610s within TSI, while CS78 does not divide them to place one around 1600s. This period is located in an interval between Spörer Minimum and Maunder Minimum. Especially, the latter peak in this period is covered by telescopic sunspot observations as discussed in previous section and by surveys for nitrate signals by McCracken (2001). Since nitrate concentration in the polar ice cores are a more direct proxy for solar flares and space weather events, we also compared the data with the nitrate data available since the mid-16th century (McCracken et al. 2001) as shown in Figure 7. Note that the ice core data may have a few years of uncertain dating, and that the usage of nitrate as an index of solar flares is controversial (Wolff et al. 2012).

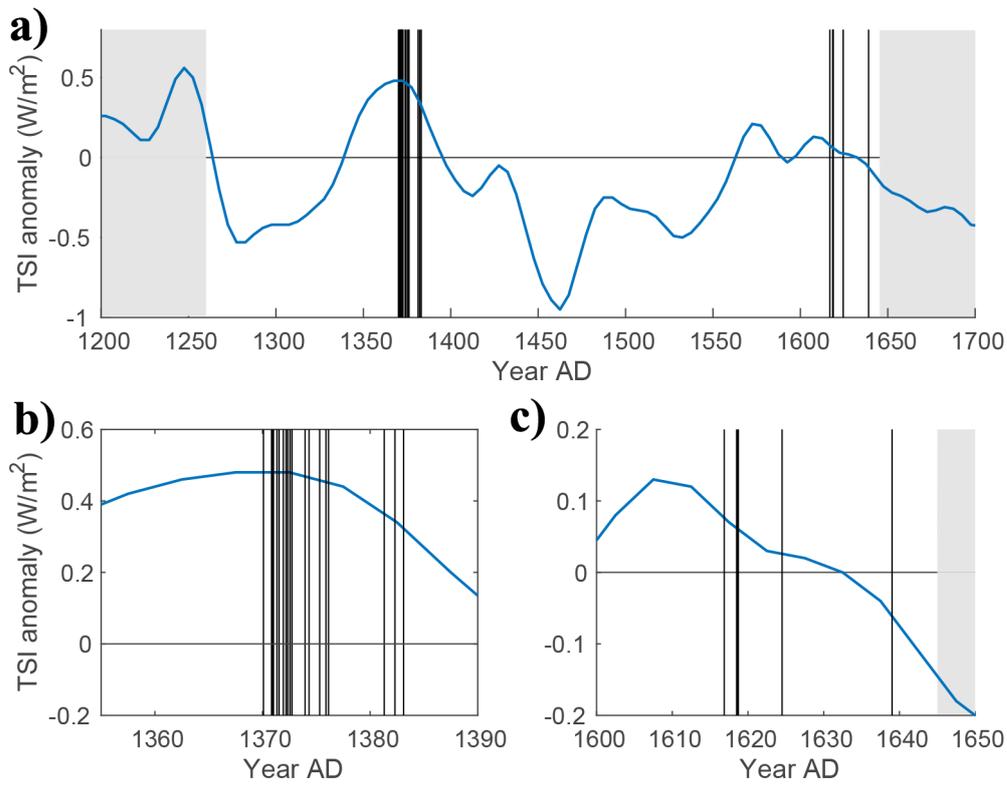

Figure 7: Records of sunspots compared with total solar irradiance (TSI) anomaly reconstructed by Steinhilber et al. (2009)

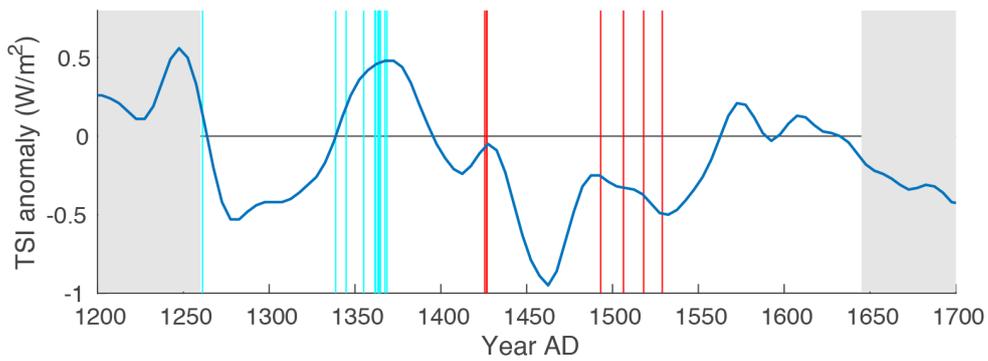

Figure 8: Records of aurora candidates compared with total solar irradiance (TSI) anomaly reconstructed by Steinhilber et al (2009)

Blue vertical lines: aurora candidates in YS

Red vertical lines: aurora candidates in MS

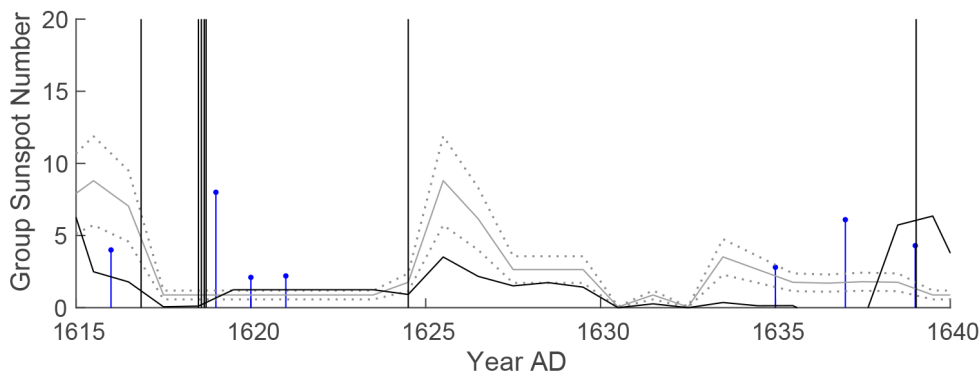

Figure 9: Comparison with group sunspot number and nitrate data

Black curve: Group sunspot number estimated by Hoyt and Schatten (1998)

Gray curve: Group sunspot numbers estimated by Svalgaard and Schatten (2016)

Gray dotted curves: Standard error limits of the estimated group sunspot number.

Black vertical lines: naked-eye sunspots

Blue vertical lines: nitrate

Figure 9 indicates that nitrate peak appear in 1619 and 1639 coinciding sunspot records in MS#S22-24 and MS#S26. Further investigations of the aurora records near these sunspot records bring us insights and evidence of intense recurrent geomagnetic storm as shown in case studies of great magnetic storms caused by intense solar flares (e.g. Gonzalez et al. 1994; Willis & Stephenson 2001; Tsurutani et al. 2003; Hayakawa et al. 2016b). For the sunspot record of MS#S23, in 1618.07, we found an aurora record on 1618.07.19 in *Qīngshǐgǎo* (ID4 in Kawamura et al. 2016).

*Qīngshǐgǎo* ID4//1618.07.19

　　天命三年⋯五月乙卯，有紅，綠，白三氣，自天下垂，覆營左右，上圓如門。（*Qīngshǐgǎo*, Astronomy XIV, 1483）

　　　　Translation: In 1618.07.19, there were three vapors whose colors were red, green, and white, suspended from the heaven, covering the encampment from left to right, as round as gates above.

Kawamura et al. (2016), calculate its normalized lunar phase as 0.93 (near new moon) and hence there is little possibility of contamination of atmospheric optics mainly caused by lunar light. Its color described "red, green, and white" is also consistent with auroras. The "red" one probably corresponds to the most common aurora emission from OI (630.0 nm). The "green" one should be the OI emission in the wavelength of 557.7 nm. The "white" one should be faint part of this aurora. As discussed in Tamazawa et al. (2017), faint auroras are frequently seen whitish due to the structure of human-eye (Purkinje effect, Purkinje 1825). This aurora candidate falls in the same month with naked eye sunspot observation of MS#23 and we can infer that the sunspot recorded as MS#23 (and by Malapert as discussed above) caused a flare and brought a magnetic storm with low latitude aurora on 1618/07/19 and scientifically detected as a nitrate signal in 1619.

　　Around the sunspot record of MS#26 on 1638.12.09, we could not find relevant aurora records in Official

Histories but one in local treatise dated 1638.12.23 that was once discussed by Willis et al. (2005), although we need to discount its reliability as discussed above and hence we do not include this record neither into our catalogue nor into any of our figures.

*Tóngchéng xùxiū xiànzhì* (桐城續修縣志)//1638.12.23[15]

　　明崇禎十一年十一月…十九日東北有赤氣數十條如劍戟排列狀（*Tóngchéng xùxiū xiànzhì*, XXIII, f. 6b)

　　　　Translation: On 1638.12.23, there were several tens bands of red vapors in north-east with their shape like swords and pikes ordered in a line.

This record is also consistent with low latitude aurora as it is observed in the direction of north-east and reflect a vertical structure similar to "swords and pikes ordered in a line" that reminds us vertical structure of intensive aurora observed *Enkoan Zuikai Zue* (猿猴庵随観図会)[16] for aurora observation on 1770.09.17 in Japan, with description of "white stripes in red vapor". We can infer that the sunspot captured as MS#26 in 1638.12.09 lasted several days and caused a flare and brought a magnetic storm with low latitude aurora on 1638.12.23 and detected as a nitrate signal in 1639 as well.

**Conclusions**

We have surveyed sunspot and aurora-like records in the Official Histories of the *Yuán* dynasty and the *Míng* dynasty during 1261-1368 and 1368-1644. We found no sunspot candidates and 20 aurora candidates during the *Yuán* period, and 10 aurora candidates and 26 sunspot candidates during the *Míng* period, with information regarding their size, color, and so on, as shown in Appendix II and Table 1. These data show us two peaks: the former during 1350s-80s and the latter during 1610s-30s. On the aspect of long-term solar activity, these peaks are consistent with contemporary TSI reconstructed by Steinhilber et al. (2009). The former one is contemporary with a series of Russian reports for huge naked-eye sunspots shown in Appendix III. The latter one is comparable with contemporary Western sunspot drawings based on telescopic observations. Further discussions allow us to compare sunspot records in the latter peak with aurora records and nitrate data. While we need to admit the possibility of contamination by the moon light within aurora candidates, the comparisons of aurora candidates with lunar phase do not show us clear influence by the moon light, partially because we cannot gain enough record numbers statistical discussions of the distribution over the lunar phases. These datasets allow us to review the solar activity before and within the earliest phase of the age of telescopic observations, both in long-term solar activity and short-term space weather events. We believe that further discussions in comparison with records from other regions allow us to reconstruct the solar activity in the said periods.

**Competing interests**

The authors declare that they have no competing interests.

---

**Author contributions**

This research was performed with the cooperation of the authors. Hisashi Hayakawa and Tadanobu Aoyama made philological and historical contributions. Harufumi Tamazawa and Akito Davis Kawamura made contributions on the interpretation and analysis of the database. Yusuke Ebihara and Hiroko Miyahara made contributions on the interpretation and analysis of the auroras. Hiroaki Isobe supervised this study. All authors read and approved the final manuscript.

**Acknowledgement**

Here we state special thanks to Dr. S. Sasaki for his helps on access and discussions on TYX, Dr. A. Tsukamoto, Prof. S. Hori, Mr. K. Saito, and Ms. M. Baba for their helpful advice on interpreting historical documents, Dr. H. Maehara, Dr. S. Honda, and Mr. Y. Notsu for discussions on superflares, Mr. T.-H.Chiu for his advices on transcriptions of classic Chinese, and National Archives of Japan for its courtesy on usage of TYX. We also acknowledge to the supports by the Center for the Promotion of Integrated Sciences (CPIS) of SOKENDAI as well as the Kyoto University's Supporting Program for Interaction-based Initiative Team Studies "Integrated study on human in space" (PI: H. Isobe), the Interdisciplinary Research Idea contest 2014 (PI: H. Isobe) and SPIRITS 2017 (PI: Y. Kano) by the Center of Promotion Interdisciplinary Education and Research, the "UCHUGAKU" project of the Unit of Synergetic Studies for Space, and the Exploratory Research Projects 2015 and 2016 (PI: H. Isobe) of the Research Institute of Sustainable Humanosphere, Kyoto University. Additionally, HH's work is supported by Grant-in-Aid for JSPS Research Fellow 17J06954 (PI: H. Hayakawa), HT and ADK's work is supported by JSPS KAKENHI Grant Number JP15H05816 (PI: S. Imada) and JP15K21709 (PI: K. Kusano), and HM's work is supported by JP15H05816 (PI: N. Yoden) and JP25287051 (PI: H. Miyahara).

**Reference**

Allen, J. 2012, Nature 486, 473.

Arlt, R., Senthamizh, P. V., Schmiel, C., & Spada, F. 2016, A&A, 595, A104.

Aulanier, G. et al. 2013, A&A, 549, 7

Baker, D. N., Li, X., Pulkkinen, A., Ngwira, C. M., Mays, M. L., Galvin, A. B., Simunac, K. D. C. 2013, Space Weather, 11, 585

Basurah, H. M. 2006, Journal of Atmospheric and Solar-Terrestrial Physics, 68, 937

Barnard, E. E. 1910, ApJ, 31, 223.

Beijing Observatory, 1985, A integrated catalogue of Chinese local treatises (Beijing: Zhinghua Book Company) [in Chinese]

Beijing Observatory, 1988, A catalogue of ancient Chinese Astronomical Catalogue (Nanjing: Phoenix Science Press). [in Chinese]

Board, S. S. 2008, Severe space weather events—understanding societal and economic impacts. (Washington DC: National Academies Press).

Boteler, D. H., Pirjola, R. J., & Nevanlinna, H. 1998, Advance in Space Research, 22, 17

Boyle, J. A. 1977, The Mongol world empire, 1206-1370 (London: Variorum Reprints).

Cade, W. B. & Chan-Park, C. 2015, Space Weather, 13, 99

Candelaresi, S. et al. 2014, ApJ, 792, 9

Carrasco, V. M. S., & Vaquero, J. M. 2016, Solar Physics, 291, 9, 2493.

Carrasco, V. M. S., Trigo, R., Vaquero, J. M. 2017, PASJ, 69, L1.

Carrington, R.C. 1859, MNRAS, 20, 13

Carrington, R.C. 1863, Observations of the spots on the sun from November 9, 1853, to March 24, 1861, made at Redhill, London.

Chamberlain, J. W. 1961, Physics of the aurora and airglow (New York: Academic Press).

Clark, B. M., Blake, C. H., Knapp, G. R. 2012, ApJ, 744, 119.

Clark, D. H., & Stephenson, F. R. 1978, QJRAS, 19: 387-410.

Clette, F. Svalgaard, L., Vaquero, J. M., Cliver, E. W. 2014, Space Science Reviews, 186, 1-4: 35-103

Cliver, E., & Svalgaard, L. 2004, Solar Physics, 224, 407.

Dell Dall'Olmo, U. 1979, JGR, 84, 1525

Davis, T. N. 1978, J. Geomagn. Geoelectr., 30, 371.

D'Ohsson, C. 1834, Histoire des Mongols, depuis Tchinguiz-Khan jusqu'à Timour Bey ou Tamerlan, I-IV, (La Haye: Van Cleef).

Eddy, J. A. 1976, Science, 192, 1189

Eddy, J. A. 1977a, Scientific American, 236, 80

Eddy, J. A. 1977b, Climatic Change, 1, 173

Gallilei, G. 1613, Istoria e dimostrazioni intorno alle macchie solari e loro accidenti. Roma: appresso Giacomo Mascardi.

Gassendi, P. 1658, Opera Omnia, (Lyon: L Anisson & JB Devenet).

Gonzalez, W. D., Joselyn, J. A., Kamide, Y., Kroehl, H. W., Rostoker, G., Tsurutani, B. T., Vasyliunas, V. M. 1994, JGR, 99, A4: 5771

Gray, L. J., et al. 2010, Rev. Geophys., 48, RG2001

Hayakawa, H., Tamazawa, H., Kawamura, A. D., Isobe, H. 2015, EPS, 67, 82

Hayakawa, H., Isobe, H., Kawamura, A. D., Tamazawa, H., Miyahara, H., Kataoka, R. 2016a, PASJ, 68, 33

Hayakawa, H., Iwahashi, K., Tamazawa, H., et al. 2016b, PASJ, 68, 99

Hayakawa, H., Mitsuma, Y., Kawamura, A. D., Ebihara, Y., Miyahara, H., Tamazawa, H., Isobe, H. 2016c, EPS, 68, 195

Hayakawa, H., Tamazawa, H., Uchuyama, Y., Ebihara, Y., Miyahara, H., Kosaka, S., Iwahashi, K., Isobe, H. 2017a, Sol. Phys., 292, 12. doi: 10.1007/s11207-016-1039-2

Hayakawa, H., Mitsuma, Y., Fujiwara, Y., et al. 2017b, PASJ, 69, 17. doi: 10.1093/pasj/psw128

Heath, A. W. 1994, J. Br. Astron. Assoc., 104, 304

Hevelius, J. 1647, Selenographia, sive Lunae description (Danzig: Andreas Hënefeld).

Ho, P-Y, & Needham, J. 1959, Weather, 14, 124

Honda, S. et al. 2015, PASJ, 67, 8510

Nogami, D. et al. 2014, PASJ, 66, L4

Hoyt, D. V., Schatten, K. H. 1998, Sol. Phys., 181, 491

Isahaya ,Y. 2010, Tarikh-e Elm: Iranian Journal for the History of Science, 8, 19.

Kaneda, E., Oguti, T., Nagata, T. 1968, Photographic atlas of auroral forms observed at Syowa station, JARE Scientific reports, Department of Polar Research, Tokyo, Ser. A., No. 4, 43.

Kawamura, A. D., Hayakawa, H., Tamazawa, H., Miyahara, H., Isobe, H. 2016, PASJ, 68, 79

Keimatsu, M. 1969a, Shirin 2, 52, 62

Keimatsu, M. 1969b, Shirin 3, 52, 14

Keimatsu, M. 1970, Annals of Science of Kanazawa University 7, 1

Keimatsu, M. 1971, Annals of Science of Kanazawa University 8, 1

Keimatsu, M. 1972, Annals of Science of Kanazawa University 9, 1

Keimatsu, M. 1973, Annals of Science of Kanazawa University 10, 1

Keimatsu, M. 1974, Annals of Science of Kanazawa University 11, 1

Keimatsu, M. 1975, Annals of Science of Kanazawa University 12, 1

Keimatsu, M. 1976, Annals of Science of Kanazawa University 13, 1

Kenyon, S. L., and Storey, J. W. V. 2006, PASP, 118, 489

Kimball, D. S. 1960, A study of the aurora of 1859, Scientific Report No. 6 (Anchorage: the University of Alaska)

Krisciunas, K. and Schaefer, B. E. 1991, PASP, 103, 1033

Kronk, G. W. 1999, Cometography: Volume 1, Ancient-1799 (New York: Cambridge University Press).

Lada, C. J. 2006, ApJ, 640, L63

Landsman, W. B. 1993, The IDL Astronomy User's Library, Astronomical Data Analysis Software and Systems II, A.S.P. Conference Series, 52, 246.

Lee, E., Ahn, Y., Yang, H., Chen, K. 2004, Sol Phys 224, 373

Loomis, E. 1860, Am J Sci 30, 339

Maehara, H., Shibayama, T., Notsu, S., Notsu, Y., Nagao, T., Kusaba, S., Honda, S., Nogami, D., Shibata, K. 2012, Nature, 485, 478

Malapert, C. 1633, Austriaca Sidera heliocyclia astronomicis hypothesibus illigata (Balthazaris Belleri).

McCracken, K. G., Dreschhoff, G. A. M., Zeller, E. J., Smart, D. F. and Shea, M. A. 2001, JGR, 106, 21, 585

Meeus, J. 1998, Astronomical algorithms. (Willmann-Bell Inc, Richmond), ISBN 0943396-61-1

Mekhaldi, F., et al. 2015, Nat. Commun., 6, 8611

Miyake, F., Nagaya, K., Masuda, K., & Nakamura, T. 2012, Nature: 486, 240

Miyake, F., Masuda, K., & Nakamura, T. 2013, Nat Commun. 4, 1748

Miyake, F., Suzuki, A., Masuda, K., et al. 2015, GRL, 42: 84-89.

Nakazawa, Y., Okada, T., Shiokawa, K. 2004, EPS, 56, e41

National Research Council 2008, Severe space weather events — understanding societal and economic impacts. National Academies Press, Washington DC

Neuhäuser, R. & Neuhäuser, D. L. 2016, Astron. Nachr., 337, 581.

Notsu, Y., Honda, S., Maehara, H., Notsu, S., Shibayama, T., Nogami, D., & Shibata, K. 2015a, PASJ, 67, 32

Notsu, Y., Honda, S., Maehara, H., Notsu, S., Shibayama, T., Nogami, D., & Shibata, K. 2015b, PASJ, 67, 33

Odenwald, S. 2015, Solar Storms: 2000 years of human calamity! Createspace Independent Publishing Platform, San Bernardino.

Oki, Y. 2009, Media Revolution in the late Ming era -- normal people started reading, (Tokyo: Tosui Shobo).

Owens, B. 2013, Nature 495, 7441, 300

Pankenier, D.W 2013, Astrology and Cosmology in Early China, Cambridge.

Peticolas, L. M., Hallinan, T. J., Stenbaek-Nielsen, H. C., Bonnell, J. W., Carlson, C. W. 2002, JGR, 107, A8

Purkinje, J. E. 1825. Neue Beiträge zur Kenntniss des Sehens in Subjectiver Hinsicht. Reimer: Berlin. pp. 109–110.

Royal Academy of Engineering (2013) Extreme space weather impacts on engineering systems and infrastructure. (London: Royal Academy of Engineering).

Rufus, W. C. 1939, Popular Astronomy 47, 5, 233–238.

Russell, C. T., et al. 2013, ApJ, 770, 38

Saito, K., & Ozawa, K. 1992, Examination of Chinese ancient astronomical records. (Tokyo: Yuzankaku Press). [in Japanese]

Sasaki, S. 2013, The Journal of Eastern Religions, 122, 24 [in Japanese]

Schäfer, B. E., King, J. R., Deliyannis, C. P. 2000, ApJ, 529, 1026-1030.

Scheiner, C. 1630, Rosa Ursina (Romae: Brassiano).

Scripta Sinica, 2014, http://hanchi.ihp.sinica.edu.tw. Accessed 3 April 2016.

Shibata, K., et al. 2013, PASJ 65:49S

Shiokawa, K., Ogawa, T., & Kamide, Y. 2005, J Geophys Res 110(9), A05202

Soderblom, D. R. 2010, ARA&A, 48, 581

Solanki, S.K., Usoskin, I.G., Kromer, B., Schüssler, B., Beer, J., 2004, Nature 431, 1084–1087.

Spuler, B. 1939, Die Mongolen in Iran: Politik, Verwaltung und Kultur der Ilchanzeit 1220-1350, (Leipzig: J.C. Hinrichs Verlag).

Spuler, B. 1943, Die Goldene Horde: die Mongolen in Rußland, 1223-1502, I-II, (Wiesbaden: O. Harrassowitz).

Steinhilber, F., Beer, J., Fröhlich, C. 2009, GRL, 36, L19704

Stephenson, F. R., Willis, D. M., Hallinan, T. J. 2004, A&G 45, 6.15

Stephenson, F. R. 2015, Advances in Space Research, 55, 1537

Sugiyama, M. 2004, The Mongolian Empire and the Great Yuán Ulus (Kyoto: Kyoto University Press). [in Japanese]

Svalgaard, L., & Schatten, K. H. 2016, Sol Phys, 291, 2653

Takahashi, T., Mizuno, Y., Shibata, K. 2016, ApJL, 833, 1, L6

Takahashi, T., Shibata, K. 2017, ApJL, in press (arxiv: 1702.06607)

Takeuchi, Y. 2002, How Chinese Official Histories have been written (Tokyo: Taishukan Shoten). [in Japanese]

Tamazawa, H., Kawamura, A. D., Hayakawa, H., Ebihara, Y., Isobe, H. 2016, PASJ, 69, 22.

Tinsley, B. A., Rohrbaugh, R. P., Rassoul, H., Barker, E. S., Cochran, A. L., Cochran, W. D., Wills, B. J., Wills, D. W., and Slater, D. 1984, GRL, 11, 572.

Tonami, M., Kishimoto, M., Sugiyama, M. 2006, An introduction to researches for Chinese history. (Nagoya: Nagoya University Press). [in Japanese]

Trondsen, T. S., Cogger, L. L. 1997, JGR, 102, 363.

Tsurutani, B. T., Gonzalez, W. D., Lakhina, G. S., & Alex, S. 2003, JGR 108, 1268

Usoskin, I. G., Solanki, S. K. & Kovaltsov, G. A. 2007, A&A, 471, 301

Usoskin, I. G. 2013, Living Review of Solar Physics 10, 1

Usoskin, I.G., Kromer, B., Ludlow, F., Beer, J., Friedrich, M., Kovaltsov, G.A., Solanki, S.K., Wacker, L. 2013 A&A, 552, L3.

Vaquero, J. M., Vazquez, M. 2009, The Sun recorded through history (Berlin: Springer).

Vesel, Z. 2002, "Islamic and Chinese Astronomy under the Mongols: a Little-Known Case of Transmission", in:

From China to Paris: 2000 Years Transmission of Mathematical Ideas (Yvonne Dold-Samplonius, Joseph W. Dauben, Menso Folkerts and Benno van Dalen, eds.), Boethius 46, 327

Vyssotsky, A. N. 1949, Astronomical records in the Russian chronicles from 1000 to 1600 AD. Meddelande Fran Lunds Astronomiska Observatorium, Ser. II, No. 126, the Observatory of Lund, Lund: Historical Notes and Papers, No. 22.

Wang, S. H., Fang, J., Chen, J. R., Zhang, J. H., Fang, Y. F. 2006, Compendium of Chinese daily calendars, IV (Changchun: Jilin history and literature press). [in Chinese]

Willis, D. M. & Stephenson, F. R. 1999, Annales Geophysicae, 18, 1

Willis, D. M. & Stephenson, F. R. 2001, Annales Geophysicae, 19, 289

Willis, D. M., Armstrong, G. M., Ault, C. E., Stephenson, F. R. 2005 Annales Geophysicae, 23, 3, 945.

Wolff, E. W., Bigler, M., Curran, M. A. J., Dibb, J. E., Frey, M. M., Legrand, M., McConnell, J. R. 2012, GRL, 39, L08503

Xu, Z., Pankenier, D. W., Jiang, Y. 2000, East Asian archaeoastronomy (Amsterdam: Gordon & Breach)

Yabuuchi, K. 1967, A history of science and technology in the Song and Yuan eras. (Kyoto: Institute for Research in Humanities of the Kyoto University) [in Japanese]

Yau, K. K. C., Stephenson, F. R. 1988, QJRAS, 29, 175

Yau, K. K. C., Stephenson, F. R., Willis, D. M. 1995, A catalogue of auroral observations from China, Korea and Japan (193 BC–AD 1770) (Chilton: Rutherford Appleton Lab.).

Zolotova, N. V. & Ponyavin, D. I. 2015, ApJ, 800, 42

## Appendix I: Black spots/vapors "near the sun (日傍/日旁)"

As explained in main text, in this paper, we surveyed black spots and black vapors in the sun (日中), and did not include those "near the sun (日傍/日旁)" that are included in some previous studies such as Xu et al. (2000) due to historical and philological view points. To complement insufficient number of examples (3 from MS as "日旁," and 1 from YS as "日傍"), here we use TYX, a contemporary manual for Chinese astronomical divination compiled by professional astronomers in *Míng* dynasty. TYX cites many examples from past documents with drawings reliably reflecting contemporary understandings of astronomy, some of which are presented in Figure 10 (a)–(c). We must note that TYX was compiled as a textbook for astrology by contemporary professional astronomers with drawings and relevant records under Emperor *Rénzōng* (仁宗, r. 1424-25) (Sasaki 2013). Their relevant text is transcribed and translated as follows:

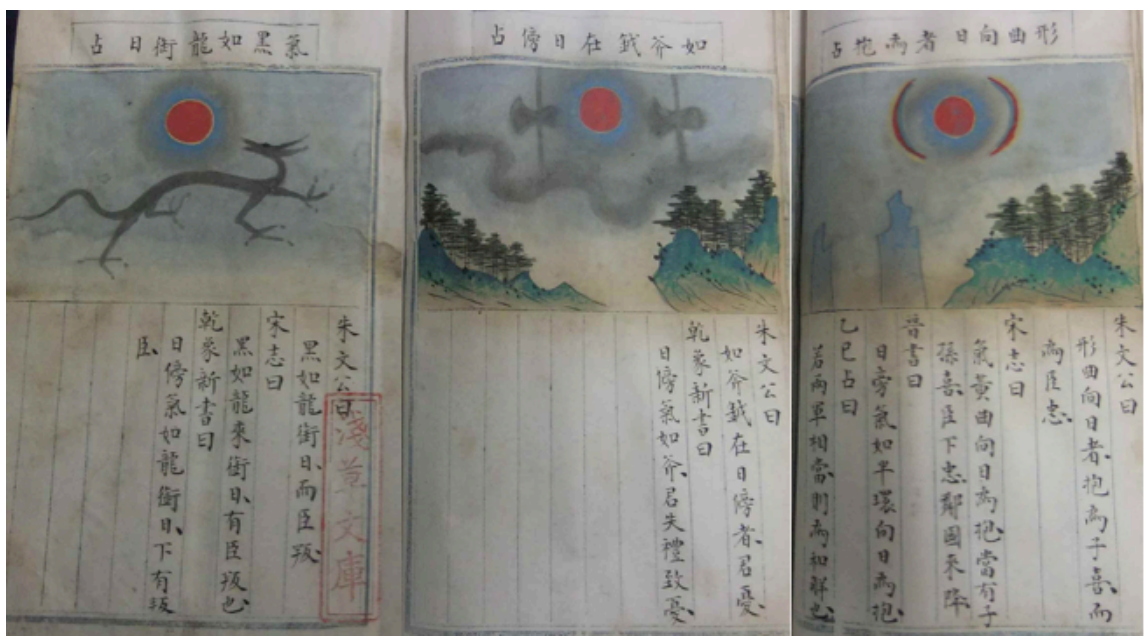

Figure 10: drawings for omens "near the sun" in TYX

(a) left: TYX: II, 1a; (b) center: TYX: II, 5b; (c) right: TYX: II, 21b.

TYX1//TYX: II, 1a, 3rd paragraph. (Figure 11a)

**Original Text:** 乾象親書曰：日傍黑氣如龍銜日，下有叛臣。

**Our Translation:** According to *Qiánxiàng Qīnshū*: If there is black vapor holding the sun in the mouth like a dragon **near the sun**, there is a revolting subordinate under the ruler.

TYX2//TYX: II, 5b (Figure 11b)

**Original Text:** 朱文公曰：如斧鉞在日傍者，君憂。乾象親書曰：日傍氣如斧鉞，君失禮致憂。

**Our Translation:** According to *Zhū Wéngōng*: If there is something like axes **near the sun**, there is grieves on the ruler. According to *Qiánxiàng Qīnshū*: If there is vapor like axes **near the sun**, the ruler is to loose the courtesy and get grieves.

TYX3//TYX: II, 21b, 3rd paragraph (Figure 11c)

**Original Text:** 晋書曰：日旁氣如半環向日為抱

**Our Translation:** According to *Jìnshū*: Near the sun, vapor like half rings head to the sun to make *bào*.

Seeing these relevant figures, we can easily understand these phenomena "near the sun (日傍/日旁)" are not in the sun but out of the sun and are hardly sunspots. Thus, black spots or black vapors "near the sun" seems not like sunspots but like halos or clouds in unusual form near the sun. Considering that TYX was a textbook of contemporary astrology with drawings compiled by contemporary professional astronomers under the Emperor's order, their drawings have sufficient reliability. Although we do not totally exclude the possibility of black spots/vapors "near the sun" as potential sunspots, these figures in TYX require us to think twice to relate them with sunspots[17].

## Appendix II: Original Text and Translations of Records of sunspots and Aurora Candidates in YS and MS

### Aurora Candidates in YS (元史)

YS#A1//1261.02.09

中統···二年春正月辛未夜，東北赤氣照人，大如席。（YS, *Shìzŭ* I, 69）

Translation: On 1261.02.09 at night, in the northeast, a red vapor shone on people as big as a seat.

〔中〕統二年正月辛未···是夜，東北有赤氣照人，大如席。（YS, Five Elements I, 1066）

Translation: On 1261.02.09 at night, in the northeast, a red vapor shone on people as big as a seat.

YS#A2//1338.08.30

至元四年八月···丁丑，白虹貫天。（YS, *Shùndì* II, 845）

Translation: On 1338.08.30, a white rainbow penetrated the heavens.

至元四年八月丁丑，京師白虹亘天。（YS, Five Elements II, 1109）

Translation: On 1338.08.30, a white rainbow extended across the heavens in *Jīngshī*.

YS#A3: 1344.09.27

至正四年八月丁丑，京師白虹亘天。（YS, Five Elements II, 1109）

Translation: On 1344.09.27, in *Jīngshī*, a white rainbow extended across the heavens.

YS#A4//1354.12.18

至正十四年···十二月辛卯，絳州北方有紅氣如火蔽天。（YS, *Shùndì* VI, 917）

Translation: On 1354.12.18, there was a scarlet vapor in the north at *Jiàngzhōu*, covering the heavens like a fire.

至正十四年···十二月辛卯，絳州有紅氣，起自北方，蔽天幾半，移時方散。（YS, Five Elements, 1102）

Translation: On 1354.12.18, a scarlet vapor appeared from the north at *Jiàngzhōu*, covering almost half of the heavens and disappearing as time passed.

YS#A5//1359.10.08（YS, *Shùndì* VIII, p949）至正十九年九月···丙午，夜，白虹貫天。

Translation: On 1359.10.08, at night, a white rainbow went across the heavens.

YS#A6//1361.08.21

---

[17] In case including black spots/vapors "near the sun," we should note one interesting record: "崇禎十二年···二月庚子，日旁有紅白丸，又白芒黑氣交掩，日光摩蕩" (MS, Astronomy III: p413). We can translate this sentence as "On 1639.03.16, there was a reddish white circle near the sun, and black vapors with white rays overwrapped one another, and the sun light was roiled and agitated." We find it difficult to relate this record with sunspot due to its location "near the sun" as explained above. However, including this record into potential sunspot records may let us relate this with an example of white light flare that is seen overwrapped with sunspots (e.g. Carrington 1859). If so, this can be earlier than the present earliest report for white light flare reported by John Flamsteed in 1705 (Carrasco & Vaquero 2016). As we find it difficult to relate black spots/vapors "near the sun" in our paper, we avoid further discussion for this record.

至正二十一年秋七月…己巳，沂州西北有赤氣蔽天如血。（YS, *Shùndì* IX, 956）

Translation: On 1361.08.21, at *Yízhōu* in the northwest, a red vapor covered the sky like blood.

至正…二十一年七月己巳，冀寧路忻州西北，有赤氣蔽空如血，逾時方散。（YS, Five Elements II, 1102）

Translation: On 1361.08.21, at *Jìnínglù Xīnzhōu* in the northwest, a red vapor covered the sky like blood and disappeared as time passed.

YS#A7//1361.09.03

至正二十一年…八月壬午，棣州夜半有赤氣亙天，起西北至于東北。（YS, Five Elements II, 1102）

Translation: On 1361.09.03, at *Dìzhōu* at midnight, a red vapor crossed the heavens appearing from the northwest to the northeast.

YS#A8//1361.09.04

至正二十一年八月…癸未，彰德西北，夜有紅氣亙天，至明方息。（YS, Five Elements II, 1102）

Translation: On 1361.09.04, at *Zhāngdé* at night, in the northwest, a scarlet vapor crossed the sky and finished at sunrise.

YS#A9//1361.09.06

至正二十一年…八月乙酉，大同路北方夜有赤氣蔽天，移時方散。（YS, *Shùndì* IX, 957）

Translation: On 1361.09.06, at *Dàtónglù* in the north, at night, a red vapor covered the heavens and disappeared as time passed.

至正二十一年八月…乙酉，大同路北方，夜有赤氣蔽天，直過天庭，自東而西，移時方散，如是者三。（YS, Five Elements II, p1102）

Translation: On 1361.09.06, at *Dàtónglù* in the north, at night, a red vapor covered the heavens. It passed by *Tiāntíng* (Leo, Com, Vir, CVn, UMa, and LMi) from East to West and disappeared as time passed. Something like this happened three times.

YS#A10//1361.11.13

至正二十一年…冬十月癸巳，絳州有赤氣見北方如火。（YS, *Shùndì* IX, 972）

Translation: On 1361.11.13, at *Jiàngzhōu*, there was a red vapor in the north like fire.

至正二十一年…十月癸巳昧爽，絳州有紅氣見于北方，如火。（YS, Five Elements II, 1103）

Translation: On 1361.11.13, in the time of twilight before the sunrise, at *Jiàngzhōu*, a scarlet vapor was observed in the north like fire.

YS#A11//1362

至正二十二年，京師有白氣如小索，起危宿，長五百丈，掃太微。（YS, Five Elements II, 1109）

Translation: In 1362, at *Jīngshī*, there was a white vapor like a rope. It appeared from *Wēisù* (Aqr & Peg) to *Tàiwēi* (Leo & Vir) as long as 500 *zhàng*.

YS#A12//1363.04.06

至正二十三年三月…壬戌，大同路夜有赤氣亙天，中侵北斗。（YS, *Shùndì* IX, 963）

Translation: On 1363.04.06, at *Dàtónglù*, at night, a red vapor crossed the heavens and came into the Plough.

至正…二十三年三月壬戌，大同路夜有赤氣亙天，中侵北斗。（YS, Five Elements II, 1103）

Translation: On 1363.04.06, at *Dàtónglù*, at night, a red vapor crossed the heavens and came into the Plough.

YS#A13//1363.07.30

至正二十三年六月…丁巳，絳州有白虹二道，衝斗牛間。（YS, *Shùndì* IX, 964）

Translation: On 1363.07.30, at *Jiàngzhōu*, there were two bands of white rainbows, hitting the area between the *Dòu* (Sgr) and the *Niú* (Cap).

至正二十三年…六月丁巳，絳州日暮有紅光見于北方，如火，中有黑氣相雜，又有白虹二，直衝北斗，逾時方散。（YS, Five Elements II, 1103）

Translation: On 1363.07.30, at *Jiàngzhōu*, at sunset, there was a scarlet light in the north, like a fire, containing black vapor and two white rainbows, which directly hit the Plough and disappeared in some time.

YS#A14//1363.08.02

至正二十三年六月…庚申，平陽路有白氣三道，一貫北極，一貫北斗，一貫天漢，至夜分乃滅。（YS, *Shùndì* IX, 964）

Translation: On 1363.08.02, in *Píngyánglù*, there were three bands of white vapor, one penetrating the North Pole, one in the Plough, and one in the Milky Way. They disappeared just at midnight.

至正二十三年六月…庚申，晉寧路北方，日暮天赤，中有白氣如虹者三，一貫北斗，一貫北極，一貫天漢，至夜分方滅。（YS, Five Elements II, 1103）

Translation: On 1363.08.02, in *Jìnnínglù*, in the north after sunset, the heaven was red; and there were three bands of white vapor like rainbow in the midst, one penetrating the Plough, one in the North Pole, and one in the Milky Way. They disappeared at midnight.

YS#A15//1363.09.27

至正二十三年八月丙辰…沂州有赤氣亙天，中有白色如蛇形，徐徐西行，至夜分乃滅。（YS, *Shùndì* IX, 964）

Translation: On 1363.09.27, in *Yízhōu*, a red vapor crossed the heavens, including white-colored vapors like the shape of serpent that went westward gradually and disappeared at midnight.

至正二十三年…八月丙辰，忻州東北，夜有赤氣亙天，中有白色如蛇形，徐徐而行，逾時方散。（YS, Five Elements II, 1103）

Translation: On 1363.09.27, in *Xīnzhōu*, in the northeast at night, a red vapor crossed the heavens, including white-colored vapors in the shape of a serpent that went gradually and disappeared as time passed.

YS#A16//1363.11.06

至正二十三年冬十月丙申朔，青齊一方赤氣千里。（YS, *Shùndì* IX, 965）

Translation: On 1363.11.06, at *Qīng* and *Qí*, there was a red vapor as long as one thousand *lǐ*;

至正二十三年…十月丙申朔，大名路向青，齊一方，有赤氣照耀千里。（YS, Five Elements II, 1103）

Translation: On 1363.11.06, at *Qīng* and *Qí* in *Dàmínglù*, there was a red vapor shining as long as one thousand *lǐ*.

**YS#A17//1364.08.09**

至正二十四年…七月癸酉, 京師赤氣滿天, 如火照人, 自寅至辰, 氣焰方息。(YS, Five Elements II, 1103)

Translation: On 1364.08.09, in *Jīngshī*, a red vapor filled up the heavens, shone on people like fire, and went from the east-northeast to the east-southeast, and, finally, the burning vapor disappeared.

**YS#A18//1364.10.08**

至正二十四年九月…癸酉, 夜, 天西北有紅光, 至東而散。(YS, *Shùndì* IX, 968)

Translation: On 1364.10.08, at night, scarlet light appeared in the northwest of the heavens and disappeared in the east.

至正…二十四年九月癸酉, 冀寧平晉縣西北方, 至夜天紅半壁, 有頃, 從東而散。(YS, Five Elements II, 1103)

Translation: On 1364.10.08, in the northwest in *Píngjìn* prefecture in *Jìníng*, at night, a scarlet heaven reddened half of heaven and disappeared in the east after a while.

**YS#A19//1367.05.29**

至正二十七年…夏五月丙子朔, 白氣二道亘天。(YS, *Shùndì* X, 978)

Translation: On 1367.05.29, two bands of white vapor crossed the heavens.

至正…二十七年五月, 大名路有白氣二道。(YS, Five Elements II, 1109)

Translation: On 1367.05.29, there were two bands of white vapor in *Dàmínglù*.

**YS#A20//1368.07.19**

至正二十八年…秋七月癸酉, 京城紅氣滿空, 如火照人, 自旦至辰方息。(YS, *Shùndì* X, 985)

Translation: On 1368.07.19, a scarlet vapor filled the sky in *Jīngchéng*, shone on people like fire, and went from from *Dàn* (3:00-5:00) to *Chén* (7:00-9:00).

**Sunspot Candidates in YS (元史)**

No records are available.

**Aurora Candidates in MS (明史)**

**MS#A1//1425.06.27**

洪熙元年六月庚戌, 中天有白氣, 東西竟天。(MS, Five Elements III, 489)

Translation: On 1425.06.27, in the mid-heaven, there was a white vapor filled up the heaven from east to west.

**MS#A2//1426.07.25**

宣德元年六月癸未夜, 有蒼白氣, 東西竟天。(MS, Five Elements III, 489)

Translation: On 1426.07.25, at night, there was a blue-white vapor filled up the heavens from east to west.

**MS#A3//1426.09.20**

宣德元年…八月庚辰，東南有白氣，狀如羣羊驚走。既滅，有黑氣如死蛇，頃之分為二。（MS, Five Elements III, 489）

Translation: On 1426.09.20, there was a white vapor in the southeast, as a surprised sheep runs. After its disappearance, there was a black vapor like a dead serpent that separated into two after a while.

MS#A4//1426.09.21

宣德元年八月辛巳，東南天有青氣，狀如人又手揖拜。（MS, Five Elements II, 479）

Translation: On 1426.09.21, there was a blue vapor in the southeast in the heavens, like a person bowing by crossing his hands.

MS#A5//1493.01.02

弘治五年十二月辛亥夜，東方有白氣，南北互天，去地五丈。（MS, Five Elements III, 489）

Translation: On 1493.01.02, at night, there was white vapor in the east, crossing the heavens from south to north, 5 *zhàng* from the earth.

MS#A6//1506.04.21

正德元年三月戊申夜，太原空中見紅光，如彎弓，長六七尺。旋變黃，又變白，漸長至二十餘丈，光芒互天。（MS, Five Elements III, 489）

Translation: On 1506.04.21, at night, a red light was observed in the sky above *Tàiyuán*, like a curving bow as long as 6 to 7 *chǐ*. Its color changed to yellow, and then to white. Gradually, its length stretched more than 20 *zhàng*, and its light crossed the heavens.

MS#A7//1506

正德元年…是歲，寧夏左屯衛紅氣互天，既而火作，城樓臺堡俱燼。（MS, Five Elements II, 464）

Translation: In 1506, at *Zuǒtúnwèi* in *Níngxià*, a scarlet vapor extended across the heavens. It then made a fire to make gate towers and forts fall down burning.

MS#A8//1518.01.17

正德…十二年閏十二月丁丑夜，瑞州有紅氣變白，形如曲尺，中外二黑氣，相闘者久之。（MS, Five Elements I, 456）

Translation: On 1518.01.17, at night, there was a scarlet vapor that changed to white in *Ruìzhōu*, like a square. Inside and outside, there were two black bands of vapor fighting each other.

MS#A9//1529.01/02[18]

嘉靖七年十二月望，白氣互天津。（MS, Five Elements III, 489）

Translation: In 1529.01/02, a white vapor extended across the Milky Way.

MS#A10//1618.11.16

---

[18] This lunar month corresponds with the dates between 1529/01/10 and 1529/02/08.

萬曆⋯四十六年九月乙卯，東南有白氣一道，闊尺餘，長二丈餘，東至軫，西入翼，十九日而滅。（MS, Astronomy III, 406）

Translation: On 1618.11.16, there was one band of white vapor in the southeast, as wide as (??) *chǐ* and longer than 2 *zhàng*, eastward to *Zhěn* (Crv) and westward to *Yì* (Crt & Hya). It disappeared after 19 days.

**Sunspot Candidates in MS (明史)**

MS#S1//1370.01.01

洪武二年十二月甲子，日中有黑子。（MS, Astronomy III, 411）

Translation: On 1370.01.01, there was a black spot in the sun.

MS#S2//1370.10.02

洪武⋯三年九月戊戌，⋯日中有黑子。（MS, Astronomy III, 411）

Translation: On 1370.10.02, there was a black spot in the sun.

MS#S3//1370.10.21

洪武三年⋯十月丁巳，⋯日中有黑子。（MS, Astronomy III, 411）

Translation: On 1370.10.21, there was a black spot in the sun.

MS#S4//1370.11.15

洪武三年十月⋯壬午，以正月至是月，日中屢有黑子，詔廷臣言得失。（MS, *Tàizǔ* II, 25）

Translation: On 1370.11.15 … From the first month to the present (tenth) month, black spots frequently appeared in the sun, and the emperor had the courtiers state if they are good or bad.

MS#S5//1370.12.07

洪武三年⋯十一月甲辰，⋯日中有黑子。（MS, Astronomy III, 411）

Translation: On 1370.12.07, there was a black spot in the sun.

MS#S6//1371.03.31

洪武⋯四年三月戊戌，⋯日中有黑子。（MS, Astronomy III, 411）

Translation: On 1371.03.31, there was a black spot in the sun.

MS#S7//1371.06.13-07.12

洪武四年⋯五月壬子至辛巳，⋯日中有黑子。（MS, Astronomy III, 411）

Translation: From 1371.06.13 to 1371.07.12, there was a black spot in the sun.

MS#S8//1371.11.06

洪武四年⋯九月戊寅⋯日中有黑子。（MS, Astronomy III, 411）

Translation: On 1371.11.06, there was a black spot in the sun.

MS#S9//1372.02.06

洪武…五年正月庚戌，…日中有黑子。（MS, Astronomy III, 411）
　　Translation: On 1372.02.06, there was a black spot in the sun.

MS#S10//1372.04.03
　　洪武五年…二月丁未，…日中有黑子。（MS, Astronomy III, 411）
　　Translation: On 1372.04.03, there was a black spot in the sun.

MS#S11//1372.06.19
　　洪武五年…五月甲子，…日中有黑子。（MS, Astronomy III, 411）
　　Translation: On 1372.06.19, there was a black spot in the sun.

MS#S12//1372.08.25
　　洪武五年…七月辛未，…日中有黑子。（MS, Astronomy III, 411）
　　Translation: On 1372.08.25, there was a black spot in the sun.

MS#S13//1373.11.15
　　洪武…六年十一月戊戌朔，…日中有黑子。（MS, Astronomy III, 411）
　　Translation: On 1373.11.15, there was a black spot in the sun.

MS#S14//1374.03.27-31
　　洪武…七年二月庚戌至甲寅，…日中有黑子。（MS, Astronomy III, 411）
　　Translation: From 1374.03.27 to 1374.03.31, there was a black spot in the sun.

MS#S15//1375.10.21
　　洪武八年…九月癸未，…日中有黑子。（MS, Astronomy III, 411）
　　Translation: On 1375.10.21, there was a black spot in the sun.

MS#S16//1375.03.23
　　洪武…八年二月辛亥，…日中有黑子。（MS, Astronomy III, 411）
　　Translation: On 1375.03.23, there was a black spot in the sun.

MS#S17//1376.01.19
　　洪武八年…十二月癸丑，…日中有黑子。（MS, Astronomy III, 411）
　　Translation: On 1376.01.19, there was a black spot in the sun.

MS#S18//1381.03.22-25
　　洪武…十四年二月壬午至乙酉，…日中有黑子。（MS, Astronomy III, 411）
　　Translation: From 1381.03.22 to 1381.03.25, there was a black spot in the sun.

MS#S19//1382.03.21

洪武…十五年閏二月丙戌，…日中有黑子。〔MS, Astronomy III, 411〕

Translation: On 1382.03.21, there was a black spot in the sun.

MS#S20//1383.01.10

洪武十五年…十二月辛巳，並如之。(日中有黑子) 〔MS, Astronomy III, 411〕

Translation: On 1383.01.10, like this (there was a black spot in the sun).

MS#S21//1616.10.10

萬曆…四十四年八月戊辰，日中有黑光。〔MS, Astronomy III, 412〕

Translation: On 1616.10.10, there was black light in the sun.

MS#S22//1618.05/06[19]

萬曆…四十六年閏四月，日中黑子相鬪。〔MS, *Dŏng Yīngjǔ*, 6289〕

Translation: In 1618.05/06, there were black spots fighting each other.

MS#S23//1618.06/07[20]

萬曆四十六年…五月朔，有黑日掩日，日無光。〔MS, *Dŏng Yīngjǔ*, 6289〕

Translation: In 1618.06/07, something black covered the sun, and the sun had no light.

MS#S24//1618.xx.xx[21]

萬曆…四十六年閏六月丙戌至戊子，黑氣出入日中摩盪。〔MS, Astronomy III, 412〕

Translation: On 1618.xx.xx, a black vapor came in and out of the sun roiling to and fro.

MS#S25//1624.05/06.xx

天啓四年…四月癸酉，日中黑氣摩盪。〔MS, Astronomy III, 412〕

Translation: In 1624.05-06.xx, a black vapor roiled in the sun.

MS#S26//1638.12.09

崇禎…十一年十一月癸亥，日中有黑子及黑青白氣。〔MS, Astronomy III, 413〕

Translation: On 1638.12.09, there was a black spot and black-blue-white vapor.

**Appendix III: Sunspot Records in Russian Chronicles**

**RC1: Никоновская летопись. *Полное собрание русских летописей (ПСРЛ)*, v. 11, St. Petersburg, 1897, p. 4**

---

[19] This lunar month corresponds with the dates between 1618.05.24 and 1618.06.21.

[20] This lunar month corresponds with the dates between 1618.06.22 and 1618.07.21.

[21] This record mistakenly dates itself in intercalary sixth month (corresponding with the dates between 1624.05.17 and 1618.06.15, although intercalary month in this year should be placed in the fourth month (see, Wang et al. 2006) and hence its date cannot be converted to a date in Western Calendar. Hereafter, xx means an error date that could not be converted from Chinese lunar calendar to Western Calendar.

Original Text: Того же лѣта бысть знаменіе на небеси, солнце бысть аки кровь, и по немъ мѣста чръны, и мъгла стояла съ поллѣта, и зной и жары бяху велицы, лѣсы и болота и земля горяше, и рѣки презхоша, иныа же мѣста воденыа до конца исхоша; и бысть страхъ и ужасъ на всѣхъ человѣцехъ и скорбь велиа.

Translation: During the same year there was a sign in the sky: The sun was like blood and there ware black spots on it, and the fog lasted for about half of the year. The swelter and the intense heat were so great. Forests, marshes and the ground burned, rivers ran dry, and the other marshy areas dried up completely; and there was terror, dread and great sorrow on all people.

### RC2: Сокращенный летописный свод 1495 г. *ПСРЛ*, v. 27, Moscow., 1962, p. 327

Original Text: Того же лѣта знамение бысть: во солнци черно, а само акы кроваво, и мъгла стояла с пол-лѣта.

Translation: During the same year there was a sign: There was something black in the sun, the sun itself was like blood, and the fog lasted for about half of the year.

### RC3: Западнорусская летопись по Супрасльскому списку. *ПСРЛ*, v. 17, St. Petersburg, 1907, pp. 36-37

Original Text: того ж лѣта бысть знамениє во солнци аки гвоздиє черно а мгла 2 мѣсяци стояла.

Translation: During the same year there was a sign like black nails in the sun, and the fog lasted 2 months.

### RC4: Устюжская летопись. *ПСРЛ, v. 37,* Leningrad, 1982, p. 73

Original Text: В лета 6873. Знамение бысть на небеси: около солнца черно, а само солнце аки кроваво, и мгла стояла с пол лета.

Translation: In the year 1365, there was a sign in the sky: There was something black near the sun, the sun itself was like blood, and the fog lasted for about half of the year.

### RC5: Никоновская летопись. *ПСРЛ*, v. 11, St. Petersburg, 1897, pp. 15-16

Original Text: Того же лѣта бысть знаменіе въ солнцѣ, мѣста чръны по солнцу аки гвозди, и мгла велика была, яко за едину сажень предъ собою не видѣти; и мнози человѣци лицемъ ударяхуся, разшедшеся, въ лице другъ друга, а птицы по воздуху не видяху летати, но падаху съ воздуха на землю, овіи о главы человѣкомъ ударяхуся; тако же и звѣри, не видяще, по селомъ ходяху и по градомъ, смѣшающеся съ человѣки, медвѣди, волцы, лисици и прочяа звѣри. Сухмень же бысть тогда великаі и зной и жарь многъ, яко устрашитися и въстрепетати людемъ; рѣки многи пресхоша, и езера, и болота; а лѣсы и боры горяху, и болота, высохши, горяху, и земля горяше; и бысть страхъ и трепеть на всѣхъ человѣцѣхъ.

Translation: During the same year there was a sign in the sun: There were black spots like nails on the sun, and the fog was so great that they couldn't see 1 sazhen (2.15m) ahead of themselves. Many people, wandering around, bumped heads together accidentally. Birds were flying through the air without sight, but fell to the ground from the air and struck a man on the head. Beasts were also not able to see, came in villages and towns, and mingled with people; bears, wolves, foxes, and other beasts. Then the drought was great, and the heat was intense, so that people trembled with fear, many rivers, lakes and marshes dried up, and forests, pine groves, parched marshes and the ground burned. There was great terror and fright on all people.

**RC6: Суздальская летопись.** *ПСРЛ,* v.1, 2nd ed., No. 3, St. Petersburg, 1927, p. 534

Original Text: того же лѣта бысть знаменіе въ солнци. мѣста черны, аки гвозди. а мгла стояла по рѧду съ два мѣсѧца. толь велика мгла была яко за двѣ сажени предъ собою не видѣти человѣка в лице. а птици по аиѥру не видаху лѣтати. но падаху на землю съ въздуха.

Translation: During the same year there was a sign in the sun: There were black spots like nails, and the fog was lasted for about 2 months. The fog was so great that they couldn't see the face of a man standing 2 sazhen (4.3m) ahead of themselves. Birds were flying through the air without sight, but fell to the ground from the air.

**RC7: Воскресенская летопись.** *ПСРЛ,* v. 8, St. Petersburg, 1859, p. 18

Original Text: Того же лѣта бысть знаменіе въ солнци: бяху по немь мѣста черны яко гвозди. Бысть же того лѣта и мгла велика поряду съ два мѣсяца, и не видѣти было передъ собою за двѣ сажени человѣка въ лице; птицы же по воздуху не видяху летати, но падаху на землю и по земли хожаху.

Translation: During the same year there was a sign in the sun: There were black spots like nails on it. During the same year the fog was also lasted for about 2 months and they couldn't see the face of a man standing 2 sazhen (4.3m) ahead of themselves. Birds were flying through the air without sight, but fell to the ground and were walking on the ground.

**RC8: Рогожский летописец.** *ПСРЛ,* v. 15, Petrograd, 1922, p. 97

Original Text: Того же лѣта бысть знаменіе въ солнци мѣста черные, акы гвозди, а мгла велика стояла по ряду съ два мѣсяца и толь велика мгла была, яко за двѣ сажени предъ собою не видѣти было человѣка въ лице, а птици по въздухоу не видяху лѣтати, но падаху съ воздуха на землю, ти тако по земли пѣши хожаху.

Translation: During the same year there was a sign in the sun: There were black spots like nails, and the great fog was lasted for about 2 months. The fog was so great that they couldn't see the face of a man standing 2 sazhen (4.3m) ahead of themselves. Birds were flying through the air without sight, but fell to the ground from the air and were walking on the ground.

**RC9: Симеоновская летописьт.** *ПСРЛ,* v. 18, 2nd ed., St. Petersburg, 1913, p. 111

Original Text: Того же лѣта бысть знаменіе въ солнци, мѣста черныя, аки гвозди, и мгла велика стояла по ряду съ два мѣсяца, и толь велика мгла была, яко за двѣ сажени предъ собою не видѣти было человѣка въ лице, а птица по воздуху не видяху лѣтати, но падаху съ воздуха на землю, ти тако пѣши хожаху по земли.

Translation: During the same year there was a sign in the sun: There were black spots like nails, and the great fog was lasted for about 2 months. The fog was so great that they couldn't see the face of a man standing 2 sazhen (4.3m) ahead of themselves. Birds were flying through the air without sight, but fell to the ground from the air and were walking on the ground.

**RC10: Львовская летопись.** *ПСРЛ,* v. 20, St. Petersburg, 1910, p. 193

Original Text: Того же лѣта бысть знаменіе во солнце: мѣста черныя, аки гвоздія; бысть же тогды и мгла велика 2 мѣсяца, яко за 2 сажени человѣка въ лице не видети, и птици на воздусе не видети, летаху, но падаху со воздуха на землю;

Translation: During the same year there was a sign in the sun: There were black spots like nails, and then the fog was so great that they couldn't see the face of a man standing 2 sazhen (4.3m) ahead, it lasted for 2 months, and birds were flying through the air without sight, but fell to the ground from the air.

**RC11: Ермолинская летопись.** *ПСРЛ,* **v. 23, St. Petersburg, 1910, p. 116**

Original Text: Того же лѣта бяху на солнци мѣста черны, яко гвоздия, и бысть же тогда мъгла велика съ два мѣсеца, яко за две сажени человѣка в лице не видять.

Translation: During the same year black spots like nails were on the sun, and then the fog so great that they can't see the face of a man standing 2 sazhen (4.3m) ahead and it lasted for about 2 months.

**RC12: Московский летописный свод конца XV века.** *ПСРЛ,* **v. 25, Moscow, 1949, pp. 186-187**

Original Text: Того же лѣта бысть знамение въ солнци; бяху по немъ мѣста черные, яко гвозди. Бысть же того лѣта и мьгла велика поряду съ два мѣсяца и не видѣти было перед собою за двѣ сажени человѣка в лице. Птици же по воздуху не видяху лѣтати, но падаху на землю и по земли хожаху.

Translation: During the same year there was a sign in the sun: There were black spots like nails on it. During the same year the great fog was also lasted for about 2 months and they couldn't see the face of a man standing 2 sazhen (4.3m) ahead of themselves. Birds were flying through the air without sight, but fell to the ground and were walking on the ground.

**RC13: Летопись свод 1497 г.** *ПСРЛ,* **v. 28, Moscow, 1963, p. 76**

Original Text: Того же лѣта бяху на солнцы мѣста черны, яко гвоздия; бысть же и мгла тогда велика с два мѣсяца, яко за двѣ сажени в лице человѣка не видѣти.

Translation: During the same year black spots like nails were on the sun, and then the fog so great that they can't see the face of a man standing 2 sazhen (4.3m) ahead and it lasted for about 2 months.

**RC14: Летопись свод 1518 г.** *ПСРЛ,* **v. 28, Moscow, 1963, p. 238**

Original Text: Того же лѣта бяхоу на солнци мѣста черны, яко гвоздия; бысть же и мгла тогда велика с два мѣсяца, яко за двѣ сажени человѣка в лице не видѣти.

Translation: During the same year black spots like nails were on the sun, and then the fog so great that they can't see the face of a man standing 2 sazhen (4.3m) ahead and it lasted for about 2 months.

**RC15: Владимирский летописец.** *ПСРЛ,* **v. 30, Moscow, 1965, p. 118**

Original Text: Того лѣта бысть знамение въ солнци, мѣста черные аки гвозди, а мгла велика стояла два мѣсеца, за две сажени не видѣти и птици не видяху лѣтати, но на землю подаху тогда.

Translation: During the year there was a sign in the sun: There were black spots like nails, and the great fog was lasted for 2 months and they couldn't see 2 sazhen (4.3m) ahead, and then birds were flying without sight, but fell to the ground.

**RC16: Пискаревский летописец.** *ПСРЛ,* **v. 34, Moscow, 1978, p. 117**

Original Text: Того же лета бысть знамение в солнце: яко гвозди по нем черны и мгла велика. Тогда же бысть не видети было перед собою за 2 сажени человека в лице; птице же по воздуху не видети летати, но падаху на землю.

Translation: During the same year there was a sign in the sun: There was something black like black nails on it, and the fog was great. Then they couldn't see the face of a man standing 2 sazhen (4.3m) ahead of themselves; birds were flying through the air without sight, but fell to the ground.

**Table 1: List of observatories in the *Yuán* Dynasty.**

| ID | Place | Pronunciation | N. Latitude in YS | N. Latitude | E. Longitude |
|----|-------|---------------|-------------------|-------------|--------------|
| 1 | 南海 | *Nánhǎi* | 15° | 23°07′ | 113°15′ |
| 2 | 衡嶽 | *Héngyuè* | 25° | 27°17′ | 112°41′ |
| 3 | 嶽臺 | *Yuètái* | 35° | unknown | unknown |
| 4 | 和林 | *Hélín* | 45° | 42°11′ | 102°49′ |
| 5 | 鐵勒 | *Tiělè* | 55° | unknown | unknown |
| 6 | 北海 | *Běihǎi* | 65° | unknown | unknown |
| 7 | 大都 | *Dàdū* | 40° | 39°54′ | 116°24′ |
| 8 | 上都 | *Shàngdū* | 43° | 42°14′ | 116°00′ |
| 9 | 北京 | *Běijīng* | 42° | 39°54′ | 116°24′ |
| 10 | 益都 | *Yìdū* | 37° | 36°51′ | 118°47′ |
| 11 | 登州 | *Dēngzhōu* | 38° | 37°48′ | 120°45′ |
| 12 | 高麗 | *Gāolí* | 38° | 37°59′ | 126°34′ |
| 13 | 西京 | *Xījīng* | 40° | 40°04′ | 113°18′ |
| 14 | 太原 | *Tàiyuán* | 38° | 37°52′ | 112°32′ |
| 15 | 安西府 | *Ānxīfǔ* | 34.5° | 34°20′ | 108°56′ |
| 16 | 興元 | *Xīngyuán* | 33.5° | 33°04′ | 107°01′ |
| 17 | 成都 | *Chéngdū* | 31.5° | 30°34′ | 104°03′ |
| 18 | 西涼州 | *Xīliángzhōu* | 40° | 38°56′ | 102°38′ |
| 19 | 東平 | *Dōngpíng* | 35° | 35°56′ | 116°28′ |
| 20 | 大名 | *Dàmíng* | 36° | 36°17′ | 115°08′ |
| 21 | 南京 | *Nánjīng* | 34° | 32°03′ | 118°47′ |
| 22 | 河南府陽城 | *Hénánfǔ Yángchéng* | 34° | 34°30′ | 113°06′ |
| 23 | 揚州 | *Yángzhōu* | 33° | 32°23′ | 119°24′ |
| 24 | 鄂州 | *Èzhōu* | 31.5° | 30°33′ | 114°18′ |
| 25 | 吉州 | *Jízhōu* | 26.5° | 27°06′ | 114°59′ |
| 26 | 雷州 | *Léizhōu* | 20° | 20°54′ | 110°05′ |
| 27 | 瓊州 | *Qióngzhōu* | 19° | 20°02′ | 110°11′ |

**Table 2: List of cities associated with the observation records from *Yuánshǐ* and *Mingshǐ*.**

| ID | Place | Pronunciation | N. Latitude | E. Longitude |
|---|---|---|---|---|
| Y1 | 京師 | *Jīngshī* | 39°54′ | 116°24′ |
| Y2 | 絳州 | *Jiàngzhōu* | 35°36′ | 111°13′ |
| Y3 | 沂州 | *Yízhōu* | 35°06′ | 118°21′ |
| Y4 | 冀寧路忻州 | *Jìnínglù Xīnzhōu* | 38°24′ | 112°44′ |
| Y5 | 棣州 | *Dìzhōu* | 37°38′ | 117°34′ |
| Y6 | 彰德 | *Zhāngdé* | 36°05′ | 114°23′ |
| Y7 | 大同路 | *Dàtónglù* | 40°04′ | 113°18′ |
| Y8 | 平陽路 | *Píngyánglù* | 36°05′ | 111°31′ |
| Y9 | 晉寧路 | *Jìnínglù* | 36°05′ | 111°31′ |
| Y10 | 冀寧路保德州 | *Jìnínglù B ǎ odézhōu* | 39°01′ | 111°05′ |
| Y11 | 冀寧平晉縣 | *Jìníng Píngjìnxiàn* | 37°52′ | 112°32′ |
| Y12 | 大名路 | *Dàmínglù* | 36°17′ | 115°09′ |
| Y13 | 京城 | *Jīngchéng* | 32°11′ | 119°25′ |
| M1 | 太原 | *Tàiyuán* | 37°52′ | 112°33′ |
| M2 | 寧夏左屯衞 | *Níngxià Zuǒtúnwèi* | 38°28′ | 106°17′ |
| M3 | 瑞州 | *Ruìzhōu* | 28°24′ | 115°23′ |

**Table 3: List of sunspot records from *Mingshǐ*. (no information on size and number of sunspots)**

| ID | Year | Month | Day | Description | Note |
|---|---|---|---|---|---|
| MS#S1 | 1370 | 01 | 01 | BS | |
| MS#S2 | 1370 | 10 | 02 | BS | |
| MS#S3 | 1370 | 10 | 21 | BS | |
| MS#S4 | 1370 | 11 | 15 | BS | from 1370.02 |
| MS#S5 | 1370 | 12 | 07 | BS | |
| MS#S6 | 1371 | 03 | 31 | BS | |
| MS#S7 | 1371 | 06 | 13 | BS | till 07.12 |
| MS#S8 | 1371 | 11 | 06 | BS | |
| MS#S9 | 1372 | 02 | 06 | BS | |
| MS#S10 | 1372 | 04 | 03 | BS | |
| MS#S11 | 1372 | 06 | 19 | BS | |
| MS#S12 | 1372 | 08 | 25 | BS | |
| MS#S13 | 1373 | 11 | 15 | BS | |
| MS#S14 | 1374 | 03 | 27 | BS | till 03.31 |
| MS#S15 | 1375 | 10 | 21 | BS | |
| MS#S16 | 1375 | 03 | 23 | BS | |
| MS#S17 | 1376 | 01 | 19 | BS | |
| MS#S18 | 1381 | 03 | 22 | BS | till 03.25 |
| MS#S19 | 1382 | 03 | 21 | BS | |
| MS#S20 | 1383 | 01 | 10 | BS | |
| MS#S21 | 1616 | 10 | 10 | BL | |
| MS#S22 | 1618 | 05/06 | | BS | In a month between 1618.05.24 and 1618.06.21 |
| MS#S23 | 1618 | 06/07 | | B | In a month between 1618.06.22 and 1618.07.21 |
| MS#S24 | 1618 | xx | xx | BV | till xx.xx |
| MS#S25 | 1624 | 05/06 | xx | BV | An error date between 1624.05.17 and 1624.06.15 |
| MS#S26 | 1638 | 12 | 09 | BS | |

1 **Table 4: List of aurora-like records from *Yuánshǐ*.**

| ID | Year | Month | Day | Color | Description | Direction | Length | Counts | Place | Notes | Normalized Moon Phase |
|---|---|---|---|---|---|---|---|---|---|---|---|
| YS#A1 | 1261 | 02 | 09 | R | V | en | | | | | 0.287 |
| | | | | R | V | en | | | | | |
| YS#A2 | 1338 | 08 | 30 | W | rainbow | | | | Y1:京師 | | 0.493 |
| | | | | W | rainbow | | | | Y1:京師 | | ==== |
| YS#A3 | 1344 | 09 | 27 | W | rainbow | | | | Y1:京師 | | 0.682 |
| YS#A4 | 1354 | 12 | 28 | R | V | n | | | Y2:絳州 | like fire | 0.116 |
| | | | | R | V | n | | | Y2:絳州 | like fire | |
| YS#A5 | 1359 | 10 | 08 | W | rainbow | | | | | | 0.533 |
| YS#A6 | 1361 | 08 | 21 | R | V | wn | | | Y3:沂州 | like blood | 0.685 |
| | | | | R | V | wn | | | Y4:冀寧路忻州 | like blood | |
| YS#A7 | 1361 | 09 | 03 | R | V | wn-en | | | Y5:棣州 | | 0.124 |
| YS#A8 | 1361 | 09 | 04 | R | V | wn | | | Y6:彰德 | until morning | 0.158 |
| YS#A9 | 1361 | 09 | 06 | R | V | n | | | Y7:大同路 | | 0.225 |
| | | | | R | V | n | | 3 | Y7:大同路 | | |
| YS#A10 | 1361 | 11 | 13 | R | V | n | | | Y2:絳州 | like fire | 0.512 |
| | | | | R | V | n | | | Y2:絳州 | like fire | |
| YS#A11 | 1362 | | | W | V | | 500z | | Y1:京師 | like a small rope | |
| YS#A12 | 1363 | 04 | 06 | R | V | | | | Y7:大同路 | | 0.732 |
| | | | | R | V | | | | Y7:大同路 | | |
| YS#A13 | 1363 | 07 | 30 | R | L | n | | | Y2:絳州 | like fire, from sunset | 0.654 |
| YS#A14 | 1363 | 08 | 02 | W | V | | | 3 | Y8:平陽路 | | 0.756 |

| YS#A16 | 1363 | 11 | 06 | R | V | | 1000 li | | Y12:青齊 | | 0.014 |
| | | | | R | V | | 1000 li | | Y12:大名路向青齊 | | |
| YS#A17 | 1364 | 08 | 09 | R | V | | | | Y1:京師 | from 3:00 to 9:00 | 0.380 |
| YS#A18 | 1364 | 10 | 08 | R | L | wn-e | | | | | 0.424 |
| | | | | R | | wn-e | | | Y11:冀寧平晉縣 | | |
| YS#A19 | 1367 | 05 | 29 | W | V | | | 2 | | | 0.019 |
| | | | | W | V | | | 2 | Y12:大名路 | | |
| YS#A20 | 1368 | 07 | 19 | R | V | | | | Y13:京城 | | 0.145 |

**Table 5: List of aurora-like records from *Míngshǐ*.**

| ID | Year | Month | Day | Color | Description | Direction | Length | Counts | Place | Notes | Normalized Moon phase |
|---|---|---|---|---|---|---|---|---|---|---|---|
| MS#A1 | 1425 | 06 | 27 | W | V | | | | | | 0.393 |
| MS#A2 | 1426 | 07 | 25 | BW | V | | | | | | 0.700 |
| MS#A3 | 1426 | 09 | 20 | W | V | es | | | | | 0.643 |
| MS#A4 | 1426 | 09 | 21 | B | V | es | | | | | 0.677 |
| MS#A5 | 1493 | 01 | 02 | W | V | e | | | | | 0.483 |
| MS#A6 | 1506 | 04 | 21 | R | L | | 20z*6~7c | | M1:太原 | | 0.949 |
| MS#A7 | 1506 | | | R | V | | | | M2:寧夏左屯衞 | | |
| MS#A8 | 1518 | 01 | 17 | W | V | | | | M3:瑞州 | | 0.210 |
| MS#A9 | 1529 | 01/02 | | W | V | | | | | In a month between 1529.01.10 and 02.08 | |
| MS#A10 | 1618 | 11 | 16 | W | V | es | 2z | 1 | | for 19 days, likely comet | 0.984 |

**Abbreviations used in Table 3 - 5.**

**Color of aurora candidates:** W white, R red/scarlet, B blue, Y yellow, Bk black, BW blue white

**Types of recorded phenomena:** V vapor (氣 *qì*), C cloud (雲 *yún*), L light (光 *guāng*)

**Direction:** e east, w west, s south, n north, m middle, en northeast (those abbreviations can be combined, for example wn-es means from northwest to southeast.)

**Units of Length:** c *chǐ*, z *zhàng*.